\documentclass[12pt, a4paper]{article}
\usepackage{amsmath,amssymb,amsfonts,bm}
\usepackage{multirow,multicol,tabularx,bigstrut}
\usepackage{caption}
\usepackage[english]{babel}
\usepackage{natbib}
\usepackage[colorlinks=true,linkcolor=blue,citecolor=blue,urlcolor=blue]{hyperref}
\usepackage{graphicx}
\usepackage[onehalfspacing]{setspace}
\usepackage{rotating}

\begin{document}
	
	\title{
		Bundle Choice Model with Endogenous Regressors: An Application to Soda Tax
	}
	\author{
		Tao Sun\footnote{
			Faculty of Business and Economics, University of Melbourne, Carlton, Victoria, 3053, Australia; Email: \texttt{tao.sun@student.unimelb.edu.au}.
			I acknowledge the financial support by the Australian Government Research Training Program (RTP) Scholarship and the Henry Buck Scholarship.
			I am very grateful to Liana Jacobi, Yong Song and Michelle Sovinsky for their guidance
			and support, as well as Rodney Strachan, Tomasz Wozniak, and participants of the European Seminar on Bayesian Econometrics 2023 and the 31st Australia New Zealand Econometric Study Group Meeting for helpful comments.
			This research was supported by the University of Melbourne’s Research Computing Services and the Petascale Campus Initiative.
			I would like to thank Information Resources Inc. (IRI) for making the data used in this paper available.
			All estimates and analyses based on IRI data set are conducted solely by the author and do not reflect the views or analyses of IRI.
		}
	}
	
	\maketitle
	
	\begin{center}{\textbf{Job Market Paper Status:} \\ 
			Latest Version Available at: \href{https://sites.google.com/view/taosun/research}{sites.google.com/view/taosun}
			\bigskip}\end{center}
	
	\begin{abstract}
		\noindent
		This paper proposes a Bayesian factor-augmented bundle choice model to estimate joint consumption as well as the substitutability and complementarity of multiple goods in the presence of endogenous regressors.
		The model extends the two primary treatments of endogeneity in existing bundle choice models: (1) endogenous market-level prices and (2) time-invariant unobserved individual heterogeneity.
		A Bayesian sparse factor approach is employed to capture high-dimensional error correlations that induce taste correlation and endogeneity.
		Time-varying factor loadings allow for more general individual-level and time-varying heterogeneity and endogeneity,
		while the sparsity induced by the shrinkage prior on loadings balances flexibility with parsimony.
		Applied to a soda tax in the context of complementarities, the new approach captures broader effects of the tax that were previously overlooked.
		Results suggest that a soda tax could yield additional health benefits by marginally decreasing the consumption of salty snacks along with sugary drinks, extending the health benefits beyond the reduction in sugar consumption alone.
		\\
	\end{abstract}
	
	\section{Introduction}
	
	The estimation of complementary choices addresses important economic and policy questions.
	Individuals and firms make complementary choices, such as the demand for milk and cereal \citep{lee2013direct}, firm-level computer purchases \citep{hendel1999estimating}, and library subscriptions \citep{nevo2005academic}.
	Understanding these choices not only sheds light on consumption behavior but also informs broader economic policies and strategies. \cite{liu2010complementarities} find that complementarities drive firms' bundling of cable TV, local phone, and broadband.
	\cite{ershov2024estimating} show that merger analyses ignoring complementarity may overstate consumer harm.
	In the context of a soda tax, sugar-sweetened beverages and other unhealthy foods may be complements.
	If this is the case, empirical studies that ignore these complementarities may fail to capture the tax’s broader effects on unhealthy complementary goods and underestimate its full health benefits.
	
	Standard discrete choice models typically assume that all choices are mutually exclusive, thereby ruling out joint consumption and complementarity.
	Relaxing this assumption induces the curse of dimensionality, as it requires redefining the choice set to include all possible bundles of choices.
	Consequently, most empirical studies \citep{gentzkow2007valuing,deza2015there} focus on applications involving only two or three products and impose restrictive assumptions.
	The need to account for endogenous regressors, a common issue in demand estimation, further exacerbates these dimensionality challenges.
	The few recent empirical studies \citep{iaria2020identification,wang2023testing,ershov2024estimating} are limited to addressing market-level endogeneity, such as market-level prices, or solely accounting for individual fixed effects.
	
	This paper proposes a Bayesian factor-augmented bundle choice model to estimate joint consumption, and the substitutability and complementarity of multiple goods, in the presence of endogenous regressors.
	The model addresses dimensionality challenges by employing a sparse factor approach to account for taste correlation and endogeneity arising from unobserved individual heterogeneities in preferences and other unobserved variables.
	Time-varying factor loadings allow for more general individual-level and time-varying heterogeneity and endogeneity,
	while the sparsity induced by the shrinkage prior on loadings balances flexibility and parsimony.
	A computationally efficient posterior inference algorithm is developed with a Markov chain Monte Carlo (MCMC) sampling scheme, which also enables posterior density predictions for estimating choice probabilities, price elasticities, and conducting counterfactual policy analysis.
	
	To address endogeneity, the proposed model extends the model with complementarity \citep{gentzkow2007valuing} and incorporates a Bayesian instrumental variables approach \citep{rossi2012bayesian,lopes2014bayesian}.
	It generalizes the two primary treatments of endogeneity in existing bundle choice models:
	(1) market-level endogenous regressors (e.g., prices) driven by time-varying unobserved variables (e.g., unobserved choice/product attributes and demand shocks) and addressed using market-level instruments \citep{iaria2020identification,ershov2024estimating},
	and (2) individual-level endogenous regressors induced by time-invariant unobserved individual heterogeneity (e.g., taste for quality) and addressed using a fixed-effects approach \citep{wang2023testing}.
	The proposed model addresses individual-level endogenous regressors resulting from time-varying unobserved variables using individual-level instruments.
	This generality allows for broader applicability, and the richer individual-level information aids in identification and empirical implementation.
	
	I employ a factor approach to model the high-dimensional error correlations that lead to taste correlation and endogeneity.
	Accounting for taste correlation is a central challenge when estimating complementarity \citep{gentzkow2007valuing}, as the observed joint consumption of two goods may result from either genuine complementarity or positive taste correlation.
	The factor approach provides a flexible and parsimonious method for modeling complex correlation structures—between choices (taste correlation), between choices and endogenous regressors (endogeneity), and across time periods.
	The time-varying factor loadings allow unobserved individual heterogeneity in preferences and demand shocks, which cause taste correlation and endogeneity, to vary over time.
	In contrast, existing approaches for handling taste correlation (e.g., random effects) are not readily scalable to applications involving endogeneity and individual-level, time-varying settings due to dimensionality constraints.
	
	Standard factor models require researchers to choose the number of latent factors to balance flexibility and parsimony, leading to model selection challenges.
	Inspired by advances in sparse Bayesian factor analysis, I adopt a cumulative shrinkage prior \citep{bhattacharya2011sparse,durante2017note} on the factor loadings.
	This prior introduces sparsity into the factor loading matrix by shrinking unimportant columns and elements to zero, allowing the model to automatically determine the number of effective factors and loadings.
	This approach addresses model selection issues in existing Bayesian choice models with a factor structure \citep{loaiza2022scalable,jacobi2024working} by jointly inferring both model parameters and complexity.
	Additionally, the asymmetric nature of the prior resolves the label-switching identification problem, ensuring a unique factor structure.
	
	The proposed model integrates bundle choice and factor structures within a Bayesian multinomial probit framework \citep{mcculloch1994exact},
	facilitating the use of data augmentation techniques to bypass the need for evaluating high-dimensional integrals when calculating choice probabilities.
	I develop a computationally efficient posterior inference algorithm using a Markov chain Monte Carlo (MCMC) sampling scheme.
	To improve algorithmic and computational efficiency, the model is fully vectorized, which mitigates dimensionality challenges and enhances the tractability of applications with modern choice data.
	This algorithm also enables posterior density predictions for estimating choice probabilities, price elasticities, and conducting counterfactual policy analysis.
	
	
	Monte Carlo simulation studies confirm that the proposed model accurately recovers the true parameters and provides reliable estimates of price elasticity.
	Comparisons between the proposed model and simpler benchmark models (time-invariant or exogenous) highlight the importance of controlling for price endogeneity when estimating complementarity, as well as the necessity of accounting for time-varying unobserved individual heterogeneities and demand shocks.
	Furthermore, the simulation studies demonstrate that as the number of individuals ($N$) or time periods ($T$) increases, the estimated own- and cross-price elasticities converge to their true values.
	
	I apply the proposed model to evaluate the implications of implementing a sugary drink tax (also known as a soda tax) in the context of complementarities.
	A primary rationale for introducing a soda tax is to reduce sugar consumption.
	However, concerns have emerged \citep{allcott2019should}.
	First, such a tax, if applied only to sugary soft drinks, may lead to increased consumption of alternative sugary products.
	Ignoring the potential for substitution could result in an overestimation of the tax's impact on overall sugar intake.
	Second, sugar-sweetened beverages and other unhealthy food items may be complements.
	If this is the case, existing empirical studies might underestimate the health benefits derived from a soda tax.
	Despite extensive research on substitution effects (see \cite{allcott2019should} for an extensive review), the potential complementarity between sugary beverages and other unhealthy foods remains understudied.
	This gap exists because standard demand models do not allow for the simultaneous estimation of substitutability and complementarity within a unified framework.
	
	The proposed bundle demand model overcomes the modeling challenge, allowing substitutability and complementarity between the goods.
	I estimate the substitution patterns between sugary soft drinks, diet soft drinks, carbonated water, milk drinks and salty snacks, using the IRI Marketing Data Set \citep{bronnenberg2008database}.
	The empirical results confirm that soda taxes decrease the intake of sugar from sugar-sweetened beverages, consequently leading to the intended health benefits.
	The results also suggest that the relationship between sugary soft drinks and milk drinks demonstrates independence.
	This finding implies that a soda tax would not lead consumers to substitute milk drinks for sugary soft drinks as an alternative sugar source.
	Furthermore, sugary soft drinks and salty snacks are identified as complements.
	This relationship suggests that the implementation of a soda tax could yield additional health benefits by marginally decreasing the consumption of salty snacks along with sugary drinks, thus extending the health benefits of the tax beyond merely reducing sugar consumption.
	
	There is a growing empirical literature on the estimation of demand for bundles and complementarity \cite{gentzkow2007valuing,crawford2012welfare,liu2010complementarities,ho2012use,gentzkow2014competition,grzybowski2016substitution,thomassen2017multi,crawford2018welfare,ershov2024estimating,ouyang2023semiparametric}.
	This paper is along the line of bundle choice models accounting for endogeneity \cite{iaria2020identification,wang2023testing,ershov2024estimating}.
	The sparse factor approach addresses the dimensionality challenges and accommodate more general individual-level and time-varying heterogeneity and endogeneity, and individual-level instruments.
	The factor approach provides an alternative to the demand inversion in models with a BLP structure \citep{berry1995automobile}.
	
	This paper introduces the first Bayesian bundle demand model for panel data, extending the Bayesian framework for cross-sectional data as developed by \cite{sovinsky2024working} for bundle demand and \cite{jacobi2024working} for bundle demand with endogenous choice sets.
	Panel data leads to gains in identification from repeated observations.
	
	This paper employs the Bayesian factor approach to model taste correlation and endogeneity.
	Known for its flexibility and parsimony in modeling error correlations, as in \cite{lopes2004bayesian, philipov2006factor, chib2006analysis, han2006asset, lopes2007factor, nakajima2012dynamic, zhou2014bayesian, ishihara2017portfolio, fruehwirthschnatter2023sparse, kastner2019sparse}, the factor approach has been utilized in various microeconometric contexts.
	These include estimating the causal effect of an endogenous binary treatment on correlated panel outcomes \citep{heckman2014treatment, jacobi2016bayesian, wagner2023factor}, and in modeling error (taste) correlations in the multinomial probit model \citep{loaiza2022scalable}.
	\cite{jacobi2024working} were the first to integrate this approach into a cross-sectional bundle demand model with endogenous choice sets.
	My model builds upon their work by incorporating more flexible structures and allowing for time-varying factor loadings.
	I employ the sparse factor approach \cite{bhattacharya2011sparse,durante2017note,kastner2019sparse} to balance flexibility and parsimony and addresses model selection issue.
	
	The application to sugary drink tax contributes to the burgeoning literature aiming to simulate the effects of soda taxes using behavior estimates from markets and periods without an existing soda tax \citep{harding2017effects, bonnet2013assessment, wang2015soda, allcott2019design, chernozhukov2019quantifying, dubois2020well}.
	To my knowledge, this study is the first to investigate the complementarity between sugar-sweetened beverages, that are subject to taxation, and other unhealthy foods.
	Furthermore, the proposed model uses individual-level prices, and accounts for price endogeneity, thereby providing more reliable elasticity estimates.
	
	The rest of the paper is organized as follows.
	Section \ref{sec:econometric_model} presents the Bayesian factor-augmented bundle demand model that accommodates endogenous regressors.
	Section \ref{sec:Bayesian_inference} discusses the choice of prior distributions and develops posterior inference and counterfactual density predictions for choice probability and price elasticity.
	Section \ref{sec:simulation} presents the simulation studies.
	Section \ref{sec:soda_application} evaluates the implications of implementing a sugary drink tax in the context of complementarities.
	The paper concludes in Section \ref{sec:conclusion}.

	\section{Econometric Model}
	\label{sec:econometric_model}
	
	This section presents the Bayesian factor-augmented bundle demand model that accommodates endogenous regressors.
	The model is an extension of Gentzkow's model with complementarity \citep{gentzkow2007valuing}, with three distinct features.
	First, as introduced in Subsection \ref{sec:illustrative_model}, the base model assumes standard normal idiosyncratic errors to facilitate Bayesian posterior inference.
	Second, Subsection \ref{sec:endogenous_regressors} extends the base model to include endogenous regressors, using price as an example.
	Third, instead of modeling taste correlations with time-invariant random effects, Subsection \ref{sec:factor_approach} introduces a factor approach to model error correlations induced by unobserved taste correlation and unobserved utility-price confounders.
	
	Subsection \ref{sec:general_model} extends the two-good illustration to the general model incorporating $J$ goods and $J_p$ endogenous regressors.
	It also presents the model vectorization to further illustrate (intertemporal) error correlations and facilitate posterior inference in the subsequent section.
	Subsection \ref{sec:comparison} compares three alternative modeling approaches for error correlations, and Subsection \ref{sec:identification} discusses identification issues.

	\subsection{Bundle Demand Model: A Two-Product Illustration}
	\label{sec:illustrative_model}
	
	Consider a scenario with two goods, denoted as $j=1,2$. Each individual $i=1,\ldots,N$ at time $t=1,\ldots,T$ chooses to consume at most one unit of each good. The individual's indirect utility from consuming each good, $j=1,2$, is
	\begin{equation}
		\bar{u}_{ijt}=\alpha p_{ijt}+x_{ijt}'\beta_j+\nu_{ijt}, \label{eq:simple_utility_j}
	\end{equation}
	where $p_{ijt}$ denotes the price of good $j$ faced by individual $i$ at time $t$.
	The price interacts with the price coefficient $\alpha$.
	The vector $x_{ijt}$ includes covariates associated with the individual $i$, good $j$ and time $t$.
	These covariates may contain good-specific intercepts, individual and product characteristics, market information and fixed effects.
	The parameter $\beta_j$ is a vector of good-specific preference parameters.
	The structural error, $\nu_{ijt}$, accounts for individual $i$'s good-specific unobserved tastes, which may vary over time.
	Importantly, these tastes may be correlated across the two goods.
	
	Choosing among 2 goods defines 4 choices, denoted as $\mathcal{R}=\{0,1,2,(1,2)\}$.
	$r=0$ represents the outside option of consuming none of the goods,
	and $r=(1,2)$ the choice of consuming both goods jointly.
	The utility an individual obtains from consuming bundle $r\in\mathcal{R}$ is given by
	\begin{align}
		u_{i1t}&=\bar{u}_{i1t}+\epsilon_{i1t}, \nonumber \\
		u_{i2t}&=\bar{u}_{i2t}+\epsilon_{i2t}, \nonumber \\
		u_{i(1,2)r}&=\bar{u}_{i1t}+\bar{u}_{i2t}+\Gamma_{(1,2)}+\epsilon_{i(1,2)t},
		\label{eq:simple_utility_r}
	\end{align}
	where $\epsilon_{irt}$ denote the idiosyncratic errors or \textit{i.i.d.} utility shocks.
	The utility for the outside option is normalized to contain only the error term, i.e., $u_{i0r}=\epsilon_{i0t}$, as the absolute levels of utilities are not identifiable. Let the idiosyncratic utility shocks follow the standard normal distribution, denoted as $\epsilon_{irt}\sim\mathcal{N}(0,1)$ for $r\in\mathcal{R}$. This induces a multinomial probit (MNP) model.
	
	The bundle-effects parameter $\Gamma_{(1,2)}$ represents the additional utility (or disutility) from consuming good 1 and 2 together. 
	It captures the substitutability and complementarity between the goods.
	\cite{gentzkow2007valuing} shows that two goods are complements if $\Gamma>0$, substitutes if $\Gamma<0$, or independent if $\Gamma=0$, in terms of Hicksian cross-price elasticity of demand.
	Furthermore, $\Gamma$ can also reflect shopping costs and preference for variety that lead to individuals purchasing a variety of goods in a time period \citep{iaria2020identification}.
	
	As standard random utility models, the observed choice, $y_{it}$, is then determined by the highest latent utility, as $y_{it}=\arg\max_r(u_{irt})$.
	The error distribution leads to the choice probability:
	\begin{equation*}
		\Pr(y_{it}=r)=\int_{u_{irt}>\{u_{i,-r,t}\}}dF(\epsilon).
	\end{equation*}
	Here, $u_{i,-r,t}$, with $-r\in\mathcal{R}\backslash r$, represents the latent utilities for the unchosen bundles,
	and $F(\epsilon)$ is the cumulative distribution function of the idiosyncratic errors.
	The integral’s range can be expressed in terms of $\epsilon_{irt}$ following equation \eqref{eq:simple_utility_r}.
	For example, when good 1 is chosen, i.e. $y_{it}=1$, the integral range can be written as a region within the support of the joint distribution $F(\epsilon_{i0t},\epsilon_{i1t},\epsilon_{i2t},\epsilon_{i(1,2)t})$, as shown below:
	\begin{equation*}
		\epsilon_{i1t}-\epsilon_{i0t}>-\bar{u}_{i1t}\quad\text{and }
		\epsilon_{i1t}-\epsilon_{i2t}>-\bar{u}_{i1t}+\bar{u}_{i2t}\quad\text{and }
		\epsilon_{i1t}-\epsilon_{i(1,2)t}>\bar{u}_{i2t}+\Gamma_{1,2}.
	\end{equation*}
	In practice, I employ the Bayesian data augmentation technique to avoid the need for evaluating the high-dimensional integrals above for calculating choice probabilities.
	
	The correlation between the unobserved tastes (structural errors) $\nu_{i1t}$ and $\nu_{i2t}$ plays a crucial conceptual role in identifying bundle effects $\Gamma_{(1,2)}$ and, in turn, assessing complementarity.
	The observed simultaneous use of two goods can arise from either genuine complementarity between them ($\Gamma_{(1,2)}>0$) or positive taste correlation ($\text{corr}(\nu_{i1t},\nu_{i2t})>0$).
	Distinguishing bundle effects from this (unobserved) taste correlation constitutes the central modeling and identification challenge in bunde demand models.
	As noted by \cite{gentzkow2007valuing}, failing to account for possible correlations in the indirect utilities of bundles containing overlapping products may lead to the identification of spurious bundle effects and complementarity.

	\subsection{Endogenous Regressors}
	\label{sec:endogenous_regressors}
	
	Endogeneity typically refers to situations where observed regressors are correlated with error terms, due to omitted variables or simultaneity.
	A classic example in choice models is the endogeneity of price, which arises from unobserved product characteristics \citep{berry1995automobile,nevo2000practitioner}.
	This subsection also uses endogenous price as an example to extend the model to accommodate endogenous regressors.
	
	Consider the price $p_{ijt}$ faced by individual $i$ at time $t$.
	This model allows prices to differ across individuals, which is a common occurrence in individual-level data due to variations in product baskets, membership discounts, volume discounts, bargaining, and other reasons.
	
	The endogenous price $p_{ijt}$ for each good $j=1,2$ is modeled by a ``first-stage" equation:
	\begin{equation}
		p_{ijt}={z_{ijt}^p}'\theta^p_j+\nu^p_{ijt}+\epsilon_{ijt}^p, \label{eq:simple_price}
	\end{equation}
	where $z_{ijt}^p$ contains a set of instrumental variables, which also includes the exogenous covariates in the utility equations (structural equations).
	$\theta_j^p$ is a vector of parameters to be estimated.
	The endogeneity arises from the correlation between the structural errors $\nu_{ijt}^p$ and $\nu_{ijt}$ from the reduced form and structural equations, respectively.
	This link induces a correlation between the endogenous price $p_{ijt}$ and unobserved characteristics $\nu_{ijt}$.
	Equivalently, we could consider that unobserved confounders, which determine both prices and utilities, are captured by the structural errors $\nu_{ijt}^p$ and $\nu_{ijt}$.
	This approach aligns with the Bayesian treatment of instrumental variables in linear models \citep{rossi2012bayesian}.
	Modeling and identifying these error correlations is crucial in addressing endogeneity, an issue closely linked to unobserved tastes in the base model.

	\subsection{Modeling Error Correlations Using Factor Approach} \label{sec:factor_approach}
	
	Error correlations play an important role in identifying complementarity in the presence of endogenous regressors.
	To consolidate this point, I define joint errors $\varepsilon$ as the summations of structural errors $\nu$ and idiosyncratic errors $\epsilon$ in the bundle utility equations \eqref{eq:simple_utility_r} (substituting the good utility equation \eqref{eq:simple_utility_j}), and the first-stage equations \eqref{eq:simple_price}.
	The idiosyncratic error only outside option $u_{i0t}=\epsilon_{i0t}$ is omitted.
	\begin{equation}
		\varepsilon_{it}
		=\begin{pmatrix}
			\varepsilon_{i1t} \\ \varepsilon_{i2t} \\ \varepsilon_{i(1,2)r} \\ \varepsilon^p_{i1t} \\ \varepsilon^p_{i2t}
		\end{pmatrix}
		=\begin{pmatrix}
			\nu_{i1t}+\epsilon_{i1t} \\ \nu_{i2t}+\epsilon_{i2t} \\ \nu_{i1t}+\nu_{i2t}+\epsilon_{i(1,2)t} \\ \nu^p_{i1t}+\epsilon^p_{i1t} \\ \nu^p_{i2t}+\epsilon^p_{i2t}
		\end{pmatrix}
		=I_\nu \nu_{it}+\epsilon_{it},
		\label{eq:joint_errors}
	\end{equation}
	where
	\begin{equation*}
		I_\nu=\begin{pmatrix}
			1&0&0&0 \\ 0&1&0&0 \\ 1&1&0&0 \\ 0&0&1&0 \\ 0&0&0&1
		\end{pmatrix}, \qquad
		\nu_{it}=\begin{pmatrix}
			\nu_{i1t} \\ \nu_{i2t} \\ \nu^p_{i1t} \\ \nu^p_{i2t}
		\end{pmatrix}, \qquad
		\epsilon_{it}=\begin{pmatrix}
			\epsilon_{i1t} \\ \epsilon_{i2t} \\ \epsilon_{i(1,2)t} \\ \epsilon^p_{i1t} \\ \epsilon^p_{i2t}
		\end{pmatrix}\sim\mathcal{N}(\mathbf{0},\mathbf{I}_5),
	\end{equation*}
	and $\mathbf{I}_5$ denotes an identity matrix with dimension 5.
	
	To estimate complementarity (bundle effects $\Gamma_{(1,2)}$), it is necessary to address the error correlation $\text{corr}(\nu_{i1t},\nu_{i2t})$.
	\cite{gentzkow2007valuing} and \cite{sovinsky2024working} tackle this issue using random effects.
	Their models assume that the vector of unobserved tastes is time-invariant and follows a multivariate normal distribution $(\nu_{i1},\nu_{i2},\ldots)'\sim\mathcal{N}(\mathbf{0},\Sigma)$ with the full covariance matrix $\Sigma$ capturing the correlation.
	\cite{deza2015there} models unobserved tastes using a latent class model, where the taste correlations are captured by class-specific intercepts.
	Similarly, addressing endogenous regressors involves modeling error correlations, $\text{corr}(\nu_{i1t},\nu_{i1t}^p)$ and $\text{corr}(\nu_{i2t},\nu_{i2t}^p)$.
	Standard Bayesian approaches employ a bivariate normal distribution \citep{rossi2012bayesian} or a mixture of bivariate normal distributions \citep{conley2008semi} to model errors in the first-stage and structural equations.
	
	The existing approaches largely rely on multivariate (mostly Gaussian) distributional assumptions for modeling errors.
	However, these established methods face challenges in the current context.
	First, the number of parameters in the full covariance matrix increases exponentially with the numbers of goods and endogenous regressors.
	While the models previously mentioned address error correlations stemming either from taste correlations or endogeneity, they do not tackle both simultaneously.
	In contrast, the model presented in this paper tackles error correlations due to both sources, resulting in higher-dimensional correlations.
	For instance, with $J=6$ goods and an equal number of endogenous regressors (prices), $J_p=6$, a full covariance matrix $\Sigma$ for time-invariant random effects would have 78 parameters.
	Identifying a large covariance matrix poses practical challenges.
	Moreover, most existing models are cross-sectional, e.g. \cite{sovinsky2024working}, \cite{rossi2012bayesian} and \cite{conley2008semi}.
	For panel models, such as \cite{gentzkow2007valuing} and \cite{deza2015there}, unobserved tastes remain time-invariant, i.e., $\nu_{ij}$.
	In contrast, the model presented here, formulated for panel data, allows for time-varying unobserved tastes, $\nu_{ijt}$.
	This added flexibility leads to a larger parameter space.
	Consider the same example with $J=J_p=6$, but now over $T=12$ time periods.
	A time-varying random-effects model without any specific structure would comprise 10,440 variance and covariance parameters.
	
	To address the challenges within a unified framework, I employ a factor approach to model the structural errors $\nu_{it}$, as follows:
	\begin{equation*}
		\nu_{it}= \left(\nu_{i1t},\nu_{i2t},\nu_{i1t}^p,\nu_{i2t}^p\right)'= \Lambda_t f_i,
	\end{equation*}
	where $\Lambda_t$ is a $4$-by-$L$ factor loading matrix,
	and $f_i\sim\mathcal{N}(\mathbf{0},\mathbf{I}_L)$ is an $L$-by-$1$ vector of independent factors with zero means and an identity covariance matrix of dimension $L$. The joint errors, as defined in Equation \eqref{eq:joint_errors}, now take the form $\varepsilon_{it}=I_\nu\Lambda_t f_i+\epsilon_{it}\sim\mathcal{N}(\mathbf{0},\Omega_{it})$ with the covariance decomposition
	\begin{equation}
		\Omega_{it}=I_\nu\Lambda_t\Lambda_t' I_\nu'+\mathbf{I}_5.
		\label{eq:joint_error_covariance}
	\end{equation}
	
	To illustrate how the factor structure captures key error correlations, consider a tri-factor ($L=3$) example, as follows:
	\begin{equation}
		\nu_{it} =\begin{pmatrix}
			\nu_{i1t} \\ \nu_{i2t} \\ \nu^p_{i1t} \\ \nu^p_{i2t}
		\end{pmatrix}
		=\Lambda_t f_i,\quad\text{with }
		\Lambda_t=\begin{pmatrix}
			\lambda_{11t}&\lambda_{12t}&0 \\ \lambda_{21t}&0&\lambda_{23t} \\ 0&\lambda_{32t}&0 \\ 0&0&\lambda_{43t}
		\end{pmatrix}
		\text{ and }f_i=\begin{pmatrix}f_{i1}\\f_{i2}\\f_{i3}\end{pmatrix}\sim\mathcal{N}(\mathbf{0},\mathbf{I}_3).
		\label{eq:factor_example}
	\end{equation}
	In this specification, the three latent factors account for taste correlation and unobserved confounders of price and utility for goods 1 and 2, respectively.
	The covariance matrix $\Omega_{it}$ of the joint errors $\varepsilon_{it}$, defined in Equation \eqref{eq:joint_error_covariance}, takes the following form:
	{\tiny
		\begin{equation*}
			\Omega_{it}=\begin{pmatrix}
				\lambda_{11t}^2+\lambda_{12t}^2+1 & & & & \\
				\lambda_{11t}\lambda_{21t} & \lambda_{21t}^2+\lambda_{23t}^2+1 & & & \\
				\lambda_{11t}^2+\lambda_{11t}\lambda_{21t}+\lambda_{12t}^2 & \lambda_{11t}\lambda_{21t}+\lambda_{21t}^2+\lambda_{23t}^2 & (\lambda_{11t}+\lambda_{21t})^2+\lambda_{12t}^2+\lambda_{23t}^2+1 & & \\
				\lambda_{12t}\lambda_{32t} & 0 & \lambda_{12t}\lambda_{32t} & \lambda_{32t}^2+1 \\
				0 & \lambda_{23t}\lambda_{43t} & \lambda_{23t}\lambda_{43t} & 0 & \lambda_{43t}^2+1
			\end{pmatrix}.
		\end{equation*}
	}The term $\lambda_{11t}\lambda_{21t}$ captures the correlation between unobserved tastes for good 1 and 2.
	The terms $\lambda_{12}\lambda_{p12}$ and $\lambda_{23}\lambda_{p23}$ capture the error correlations resulting from price endogeneity.
	
	The factor approach is a flexible and parsimonious method for modeling correlations resulting from endogeneity, between choices and across time periods.
	In the above example, a time-invariant random-effects model with a full covariance matrix for the structural errors $\nu_{it}$ would require 10 parameters.
	While, the factor structure models these correlations more parsimoniously with 6 parameters.
	This method finds application in various models, such as estimating the causal effect of an endogenous binary treatment on correlated panel outcomes \citep{jacobi2016bayesian,wagner2023factor}, and in modeling error (taste) correlations in multinomial probit model \citep{loaiza2022scalable}.
	\cite{jacobi2024working} introduce the factor approach into a cross-sectional bundle demand model with choice set endogeneity.
	This paper extends their approach to a panel data setting with more general and time-varying factor loading matrices, and applies it to models with endogenous regressors.

	
	\subsection{Identification}
	\label{sec:identification}
	
	
	This subsection discusses the identification of the exogenous bundle demand model, the extended components of endogenous regressor, and the augmented factor loadings in sequence, in the context of 2 goods.
	
	Identification of the bundle demand model involves the mean utility parameters $\theta_1$ and $\theta_2$, bundle effects $\Gamma_{(1,2)}$, and the error distribution $F(\nu_{i1t},\nu_{i2t})$.
	The mean utility parameters $\theta_1$ and $\theta_2$ are identified by the observed choice probabilities $\Pr(y_{it}=1)$ and $\Pr(y_{it}=2)$.
	The key challenge lies in separately identifying the bundle effects $\Gamma_{(1,2)}$ and the error correlation $\mathrm{corr}(\nu_{i1t}, \nu_{i2t})$.
	If goods 1 and 2 are frequently consumed together (high $\Pr(y_{it}=(1,2))$), this can be attributed to either a high value of $\Gamma_{(1,2)}$ (indicating complementarity) or a high value of $\mathrm{corr}(\nu_{i1t}, \nu_{i2t})$ (indicating positively correlated tastes for the goods).
	
	One source of identification is exclusion restrictions.
	That is, there exists a variable, such as price, that enters the utility of one product, $\bar{u}_{ijt}$, but not the other, $\bar{u}_{i,-j,t}$.
	As argued by \cite{gentzkow2007valuing}, if the goods are complements ($\Gamma_{(1,2)}>0$), shifting up the utility of good 1 (e.g., by decreasing its price $p_1$) should increase the probability of consuming good 2.
	In contrast, if $\Gamma_{(1,2)}=0$, the probability of consuming good 2 should remain unchanged.
	Formalizing Gentzkow's identification argument, \cite{fox2017note} prove that an excluded variable for each product utility $\bar{u}_{ijt}$ with large support (i.e., spanning the real line $\mathbb{R}$) provides the separate identification of the bundle effects $\Gamma{(1,2)}$ and the joint error distribution $F(\nu_{i1t},\nu_{i2t})$.
	Additionally, \cite{keane1992note} provides evidence on the role of exclusion restrictions in identifying the covariance parameters in multinomial probit models through Monte Carlo tests and an application to real-world data.
	
	Identification of the bundle demand model with endogeneity builds upon the argument presented above.
	The identification relies on excluded variables, or instruments, as in instrumental variables models.
	\cite{lewbel2000semiparametric} develops identification and estimation of binary and multinomial choice models with endogeneity, leveraging the special regressor assumption.
	Under this assumption, each utility and first-stage equation requires a special regressor with large support (real line $\mathbb{R}$) that is excluded from other equations.
	As a side note, \cite{fox2017note} employs a similar assumption.
	
	The identification of the factor-augmented model with endogeneity involves two steps: (1) identifying model parameters, $\theta$ and $\Gamma$, and error distribution, $F(\nu)$, and (2) recovering the factor loadings $\Lambda$ from $F(\nu)$.
	Meanwhile, in most cases, since the estimation of the actual factor loadings is not the primary concern but rather than a means to estimate and predict the covariance structure of $F(\nu)$, a unique identification of the factor loadings $\Lambda$ is not necessary. This allows leaving the factor loading matrix completely unrestricted.

	

	\subsection{General Model with Vectorization}
	\label{sec:general_model}
	
	This subsection presents the general model incorporating $J$ goods and $J_p$ endogenous regressors.
	The model is then vectorized to further illustrate (intertemporal) error correlations and facilitate posterior inference in the subsequent section.
	
	Index goods by $j\in\mathcal{J}=\{1,\ldots,J\}$.
	Each individual $i\in\mathcal{N}=\{1,\ldots,N\}$ at time $t\in\mathcal{T}_i=\{1,\ldots,T_i\}$ chooses to consume a bundle (subset) of the goods, i.e., $r\subseteq\mathcal{J}$.
	Denote all time periods by $\mathcal{T}=\bigcup_{i\in\mathcal{N}}\mathcal{T}_i$ and the complete choice set by $\mathcal{R}=\{r\subseteq\mathcal{J}\}$.
	Note that the outside option (consuming none of $\mathcal{J}$) is in the choice set, i.e., $r=\varnothing\in\mathcal{R}$.
	As a matter of convention, I relabel the outside option as $r=0$.
	
	Restate the indirect utilities for individual $i$ from consuming a single good $j$ at time $t$ (structural equation) and the first-stage equations for endogenous regressors $p_{ijt}$ in $z_{ijt}$, as defined in Equations \eqref{eq:simple_utility_j} and \eqref{eq:simple_price},
	\begin{align}
		\bar{u}_{ijt}&=z_{ijt}'\theta_j+\nu_{ijt}, && \text{for } j\in\mathcal{J}, \label{eq:general_utility_j} \\
		p_{ijt}&={z_{ijt}^p}'\theta^p_j+\nu^p_{ijt}+\epsilon_{ijt}^p, && \text{for } j\in\mathcal{J}_p, \label{eq:general_price}
	\end{align}
	where vector $z_{ijt}$ collects the endogenous and exogenous regressors that enter good utility equation, $\theta_j$ is a vector of good-specific preference parameters,
	and $\mathcal{J}_p$ denotes the set of endogenous regressors.
	
	Similar to Equation \eqref{eq:simple_utility_r}, define the utility of individual $i$ from consuming bundle $r$ at time $t$ as
	\begin{equation}
		u_{irt}=\sum_{j\in r}\bar{u}_{ijt}+\sum_{j_1,j_2\in r}w_{i(j_1,j_2)t}'\gamma_{(j_1,j_2)}+\epsilon_{irt},
		\label{eq:general_utility_r}
	\end{equation}
	for inside options $r\neq0$, and $u_{i0t}=\epsilon_{i0t}$ for the outside option.
	The first term on the right-hand side sums the utilities derived from each good $j$ in the bundle $r$.
	The second term, bundle effects or demand synergies, $\Gamma_{irt}=\sum_{j_1,j_2\in r}w_{i(j_1,j_2)t}'\gamma_{(j_1,j_2)}$ captures the extra utility or disutility from consuming each pair of goods $j_1$ and $j_2$ jointly in bundle $r$.\footnote{\cite{gentzkow2007valuing} finds that allowing tri-good bundle effects barely changes the results. Besides, multi-good bundle effects lead to the curse of dimensionality in the number of goods.}
	The bundle effects are modeled as a function of observable characteristics (and an intercept) $w_{i(j_1,j_2)t}$.
	The idiosyncratic utility shocks follow the standard normal distribution, denoted as $\epsilon_{irt}\sim\mathcal{N}(0,1)$ for $r\in\mathcal{R}$, inducing a multinomial probit (MNP) model.
	The observed choice, $y_{it}$, is determined by the highest latent utility, $y_{it}=\arg\max_{r\in\mathcal{R}}(u_{irt})$.
	
	I highlight model features that extend the illustrative model presented in Subsection \ref{sec:illustrative_model}.
	First, the general model incorporates $J$ goods with $J_p=|\mathcal{J}_p|$ endogenous regressors. As the number of (inside) options $R=2^J-1$ increases exponentially with $J$, it is recommended in practice to limit $J\leq7$.
	To accommodate a larger number of goods, one may consider limiting the choice set to single-good and two-good bundles, as in \cite{iaria2020identification}.
	Additionally, $J_p$ does not necessarily equal to $J$; each good's utility, $\bar{u}_{ijt}$, can include a different number of endogenous regressors.
	Second, bundle effects, $\Gamma_{irt}$, are now time-varying and heterogeneous across individuals, as a function of observable characteristics $w_{i(j_1,j_2)t}$.
	Third, the model allows for common coefficients across regressors in different equations.
	For example, the same price coefficient $\alpha$ interacts with prices in all utility equations $\alpha p_{ijt}$ for all $j\in\mathcal{J}$.
	This homogeneous-effects restriction can be similarly applied to other regressors in $z_{ijt}$, $z^p_{ijt}$ and $w_{i(j_1,j_2)t}$.
	Lastly, the model allows sparsity in the factor loading matrix $\Lambda_t$ implied by prior knowledge or economic theory, as in example \eqref{eq:factor_example}.
	
	Next, I define the vectorized model. Substitute good utility \eqref{eq:general_utility_j} into bundle utility \eqref{eq:general_utility_r}:
	\begin{equation}
		u_{irt}=\sum_{j\in r}z_{ijt}'\theta_j+\sum_{j_1,j_2\in r}w_{i(j_1,j_2)t}'\gamma_{(j_1,j_2)}+\sum_{j\in r}\nu_{ijt}+\epsilon_{irt}.
		\label{eq:general_utility_r_2}
	\end{equation}
	Stack bundle utilities over inside options $r\in\mathcal{R}\backslash0$, and first-stage equations over $j\in\mathcal{J}_p$:
	\begin{align}
		u_{it}&=z_{it}\theta+w_{it}\gamma+I_\nu^u\nu_{it}^u+\epsilon_{it}^u, &&\text{ for }t\in\mathcal{T}_i,i\in\mathcal{N}, \label{eq:utility_it} \\
		p_{it}&=z_{it}^p\theta^p+I_\nu^p\nu_{it}^p+\epsilon_{it}^p, &&\text{ for }t\in\mathcal{T}_i,i\in\mathcal{N}, \label{eq:price_it}
	\end{align}
	where parameter vectors are defined as $\theta=(\theta_1',\ldots,\theta_J')'$, $\gamma=(\gamma_{(1,2)}',\ldots,\gamma_{(J-1,J)}')'$, $\theta^p=({\theta^p_1}',\ldots,{\theta^p_{J_p}}')'$. Similarly, structural error vectors are $\nu^u_{it}=(\nu_{i1t},\ldots,\nu_{iJt})'$ and $\nu^p_{it}=(\nu^p_{i1t},\ldots,\nu^p_{iJ_pt})'$. Further stack the latent utilities and endogenous regressors $y_{it}^*=(u_{it}',p_{it}')'$:
	\begin{align}
		y_{it}^*&=h_{it}\Theta+I_\nu\nu_{it}+\epsilon_{it}, \label{eq:y_star_RE} \\
		&=h_{it}\Theta+I_\nu\Lambda_t f_i+\epsilon_{it}, \qquad\text{ for }t\in\mathcal{T}_i,i\in\mathcal{N}, \label{eq:y_star_FA}
	\end{align}
	where $\Theta=(\theta',\gamma',{\theta^p}')'$ and $\nu_{it}=({\nu_{it}^u}',{\nu_{it}^p}')'=\Lambda_t f_i$ augmented with the factor structure. Lastly, stack over $t\in\mathcal{T}_i$ for each individual $i$, and then over $i\in\mathcal{N}$:
	\begin{align}
		y_i^*&=h_i\Theta+I_{\nu,i}^T\Lambda f_i+\epsilon_i, \qquad\text{ for }i\in\mathcal{N}, \label{eq:y_star_i} \\
		y^*&=H_{zw}\Theta+H_\Lambda f+\epsilon, \label{eq:y_star_1} \\
		&=H_{zw}\Theta+H_f\lambda+\epsilon. \label{eq:y_star_2}
	\end{align}
	Details on model vectorization, along with an example with $J=3$ goods and $J_p=3$ endogenous regressors, are provided in Appendix \ref{sec:appendix_vectorization}.
	The vectorized model promotes the forthcoming comparison of alternative modeling approaches for error correlations, defines more compact notations, and facilitates an efficient Markov-Chain Monte Carlo (MCMC) sampling scheme for posterior inference.

	\subsection{Alternative Approaches for Error Correlations}
	\label{sec:comparison}
	
	This subsection compares three alternative modeling approaches for error correlations.
	They differ in the distributional assumptions imposed on the unobserved tastes and utility-price confounders (structural errors), $\nu_{it}=(\nu_{i1t},\ldots,\nu_{iJt},\nu^p_{i1t},\ldots,\nu^p_{iJ_pt})$.
	\begin{itemize}
		\item[1.] \textbf{RE} Random effects with time-invariant unobserved tastes and utility-price confounders:
		\begin{equation*}
			\nu_{it}=\nu_{i}\sim\mathcal{N}(\mathbf{0},\Sigma),\qquad \Sigma=\begin{pmatrix}    \bm{\sigma}_{j_1,j_2} & \bm{\sigma}_{j_1,J+j_2} \\ \bm{\sigma}_{j_1,J+j_2}' & \bm{\sigma}_{J+j_1,J+j_2}\end{pmatrix},
		\end{equation*}
		where full covariance matrix $\Sigma$ contains 3 variance-covariance matrices.
		$\bm{\sigma}_{j_1,j_2}$, for $j_1,j_2\in\mathcal{J}$, (a $J$-by-$J$ matrix) corresponds to the good utility equations $\bar{u}_{ijt}$, and governs the utility variances and taste correlations.
		$\bm{\sigma}_{j_1,J+j_2}$, for $j_1\in\mathcal{J}$ and $j_2\in\mathcal{J}_p$, (a $J$-by-$J_p$ matrix) accounts for the correlations between utilities and endogenous regressors, thereby capturing endogeneity.
		$\bm{\sigma}_{J+j_1,J+j_2}$, for $j_1,j_2\in\mathcal{J}_p$, (a $J_p$-by-$J_p$ matrix) corresponds to the first-stage equations.
		
		\item[2.] \textbf{FA} Factor-augmented unobserved tastes and utility-price confounders with time-invariant factor loadings:
		\begin{equation*}
			\nu_{it}=\nu_{i}=\Lambda f_i,\qquad
			\Lambda=\begin{pmatrix} \bm{\lambda}_{1} & \cdots & \bm{\lambda}_{l} & \cdots & \bm{\lambda}_{L} \\ \bm{\lambda}^p_{1} & \cdots & \bm{\lambda}^p_{l} & \cdots & \bm{\lambda}^p_{L} \\ \end{pmatrix},\qquad\text{and }f_i\sim\mathcal{N}(\mathbf{0},\mathbf{I}_L),
		\end{equation*}
		where, for $l=1,\ldots,L$, factor loadings $\bm{\lambda}_{l}=(\lambda_{1l},\ldots,\lambda_{Jl})'$ enter the good utility equations $\bar{u}_{ijt}$, and $\bm{\lambda}^p_{l}=(\lambda^p_{1l},\ldots,\lambda^p_{J_pl})'$ enter first-stage equations $p_{ijt}$.
		
		\item[3.] \textbf{TV-FA} Time-varying factor-augmented unobserved tastes and utility-price confounders:
		\begin{equation*}
			\nu_{it}=\Lambda_t f_i,\qquad
			\Lambda_t=\begin{pmatrix} \bm{\lambda}_{1t} & \cdots & \bm{\lambda}_{lt} & \cdots & \bm{\lambda}_{Lt} \\ \bm{\lambda}^p_{1t} & \cdots & \bm{\lambda}^p_{lt} & \cdots & \bm{\lambda}^p_{Lt} \\ \end{pmatrix},\qquad\text{and }f_i\sim\mathcal{N}(\mathbf{0},\mathbf{I}_L),
		\end{equation*}
		where, for $l=1,\ldots,L$, factor loadings $\bm{\lambda}_{lt}=(\lambda_{1lt},\ldots,\lambda_{Jlt})'$ enter the good utility equations $\bar{u}_{ijt}$, and $\bm{\lambda}^p_{lt}=(\lambda^p_{1lt},\ldots,\lambda^p_{J_plt})'$ enter first-stage equations $p_{ijt}$.
	\end{itemize}
	Both the RE and FA approaches restrict unobserved tastes and utility-price confounders, denoted as $\nu_i$, to be time-invariant.
	Consequently, they are inadequate for capturing temporal demand shocks that impact both utilities and market equilibrium prices.
	Similarly, these approaches fall short in accounting for temporal utility shocks that influence preferences for multiple goods.
	
	Furthermore, the random-effects approach cannot be readily extended to account for time-varying $\nu_{it}$ due to dimensionality challenges.
	A time-varying random effects would be defined on the vector $\nu_{i}=(\nu_{i1},\ldots,\nu_{iT})'\sim\mathcal{N}(\mathbf{0},\Sigma)$.
	Consider an application with $J=6$ goods and an equal number of endogenous regressors $J_p=6$, across $T=12$ periods.
	The full covariance matrix $\Sigma$ would then have $(J+J_p)T\times((J+J_p)T+1)/2=10,440$ free parameters.
	
	To illustrate how the three approaches capture error correlations induced by unobserved tastes and utility-price confounders, one can examine the covariance matrix $\Omega_i$ of the joint error vector $\varepsilon_i$.
	The process of model vectorization defines joint error vectors at different levels of vectorization.
	In Equations \eqref{eq:y_star_RE}, \eqref{eq:y_star_FA}, for $t\in\mathcal{T}_i$ and $i\in\mathcal{N}$,
	\begin{align*}
		\textbf{RE}& &\varepsilon_{it} &=I_\nu\nu_i+\epsilon_{it}\sim\mathcal{N}(\mathbf{0},\Omega_{it}),\quad &\Omega_{it} &=\mathrm{cov}(\varepsilon_{it})=I_\nu\Sigma I_\nu'+\mathbf{I}, \\
		\textbf{FA}& &\varepsilon_{it} &=I_\nu\Lambda f_i+\epsilon_{it}\sim\mathcal{N}(\mathbf{0},\Omega_{it}), \quad &\Omega_{it} &=\mathrm{cov}(\varepsilon_{it})=I_\nu\Lambda\Lambda'I_\nu'+\mathbf{I}, \\
		\textbf{TV-FA}& &\varepsilon_{it} &=I_\nu\Lambda_t f_i+\epsilon_{it}\sim\mathcal{N}(\mathbf{0},\Omega_{it}), \quad &\Omega_{it} &=\mathrm{cov}(\varepsilon_{it})=I_\nu\Lambda_t\Lambda_t'I_\nu'+\mathbf{I}.
	\end{align*}
	The covariance parameters in $\Omega_{it}$ capture error correlations between latent utilities (taste correlations) and between utility and endogenous regressors (endogeneity) within each time period $t\in\mathcal{T}$.
	
	In Equation \eqref{eq:y_star_i} and its RE and time-invariant FA counterparts, for $i\in\mathcal{N}$,
	\begin{align}
		\textbf{RE}& &\varepsilon_i &=I_{\nu,i}\nu_i+\epsilon_i\sim\mathcal{N}(\mathbf{0},\Omega_i), \quad &\Omega_{i} &=\mathrm{cov}(\varepsilon_{i})=I_{\nu,i}\Sigma I_{\nu,i}'+\mathbf{I}, \nonumber\\
		\textbf{FA}& &\varepsilon_i &=I_{\nu,i}\Lambda f_i+\epsilon_i\sim\mathcal{N}(\mathbf{0},\Omega_i), \quad &\Omega_{i} &=\mathrm{cov}(\varepsilon_{i})=I_{\nu,i}\Lambda\Lambda'I_{\nu,i}'+\mathbf{I}, \nonumber\\
		\textbf{TV-FA}& &\varepsilon_i &=I_{\nu,i}^T\Lambda f_i+\epsilon_i\sim\mathcal{N}(\mathbf{0},\Omega_i), \quad &\Omega_{i} &=\mathrm{cov}(\varepsilon_{i})=I_{\nu,i}^T\Lambda\Lambda'{I_{\nu,i}^T}'+\mathbf{I}. \label{eq:joint_error}
	\end{align}
	Covariance matrix $\Omega_{i}$ further captures intertemporal error correlations.
	
	Table \ref{tab:model_Omega} summarizes the key variances and covariances in the joint error covariance matrix $\Omega_i$ across models.
	It highlights the differences in how the RE (Random Effects), FA (Factor-Augmented), and TV-FA (Time-Varying Factor-Augmented) approaches capture error correlations induced by unobserved tastes and utility-price confounders.
	Notably, the intertemporal correlations are fixed for both the RE and FA models.
	
	\begin{table}[!h]
		\centering
		\caption{Key Elements in Joint Error Covariance Matrix $\Omega_i$ Across Models}
		\resizebox{\textwidth}{!}{
			\footnotesize
			\begin{tabular}{lccccc}
\hline
Description & Variance/Covariance & RE    & FA    & TV-FA \bigstrut\\
\hline
Utility Error Variance & $\mathrm{var}(\varepsilon_{ijt})$ & $1+\sigma_{j,j}$ & $1+\sum_{l=1}^L\lambda_{jl}^2$ & $1+\sum_{l=1}^L\lambda_{jlt}^2$ \bigstrut[t]\\
Unobserved Tastes & $\mathrm{cov}(\varepsilon_{ij_1t},\varepsilon_{ij_2t})$ & $\sigma_{j_1,j_2}$ & $\sum_{l=1}^L\lambda_{j_1l}\lambda_{j_2l}$ & $\sum_{l=1}^L\lambda_{j_1lt}\lambda_{j_2lt}$ \\
Intertemporal Tastes Corr. & $\mathrm{cov}(\varepsilon_{ijt_1},\varepsilon_{ijt_2})$ & $\sigma_{j,j}$ & $\sum_{l=1}^L\lambda_{jl}^2$ & $\sum_{l=1}^L\lambda_{jlt_1}\lambda_{jlt_2}$ \\
Regressor Endogeneity & $\mathrm{cov}(\varepsilon_{ij_1t},\varepsilon^p_{ij_2t})$ & $\sigma_{j_1,J+j_2}$ & $\sum_{l=1}^L\lambda_{j_1l}\lambda^p_{j_2l}$ & $\sum_{l=1}^L\lambda_{j_1lt}\lambda^p_{j_2lt}$ \\
Reduced Form Eq. Error Var. & $\mathrm{var}(\varepsilon^p_{ijt})$ & $1+\sigma_{J+j,J+j}$ & $1+\sum_{l=1}^L(\lambda^p_{jl})^2$ & $1+\sum_{l=1}^L(\lambda^p_{jlt})^2$ \\
ER Error Correlation & $\mathrm{cov}(\varepsilon^p_{ij_1t},\varepsilon^p_{ij_2t})$ & $\sigma_{J+j_1,J+j_2}$ & $\sum_{l=1}^L\lambda^p_{j_1l}\lambda^p_{j_2l}$ & $\sum_{l=1}^L\lambda^p_{j_1lt}\lambda^p_{j_2lt}$ \\
ER Intertemporal Corr. & $\mathrm{cov}(\varepsilon^p_{ijt_1},\varepsilon^p_{ijt_2})$ & $\sigma_{J+j,J+j}$ & $\sum_{l=1}^L(\lambda^p_{jl})^2$ & $\sum_{l=1}^L\lambda^p_{jlt_1}\lambda^p_{jlt_2}$ \bigstrut[b]\\
\hline
\multicolumn{6}{p{\textwidth}}{Notes:
$\Omega_i=\mathrm{cov}(\varepsilon_i)$, with Equation \eqref{eq:joint_error} defining $\varepsilon_i$.
$\varepsilon_{ijt}=\nu_{ijt}+\epsilon_{ijt}=\sum_{l=1}^L\lambda_{jlt}f_{il}+\epsilon_{ijt}$ corresponds to the single good bundles, $r=j$,
and $\varepsilon^p_{ijt}=\nu^p_{ijt}+\epsilon^p_{ijt}=\sum_{l=1}^L\lambda^p_{jlt}f_{il}+\epsilon^p_{ijt}$ corresponds to the first-stage equations for $j\in\mathcal{J}_p$.
ER stands for ``endogenous regressor", which refers to the first-stage equations. RE, FA, and TV-FA represent the random-effects, factor-augmented, and time-varying factor-augmented bundle demand models, respectively.
} \bigstrut[t]\\
\end{tabular}

}
		\label{tab:model_Omega}
	\end{table}

	\section{Bayesian Inference}
	\label{sec:Bayesian_inference}
	
	This section discusses the selection of prior distributions and the sampling scheme for the three groups of model parameters: augmented latent utilities $u$, equation parameters $\Theta$, and factors and loadings $\lambda$ and $f$.
	
	The joint posterior distribution of the model parameters, conditional on observed data (bundle choices $y$ and endogenous regressors $p$) can be written as a product of the likelihood and the prior, $p(u,\Theta,\lambda,f)\propto p(y,u,p|\Theta,\lambda,f)\times p(\Theta)p(\lambda)p(f)$.
	The likelihood component, augmented with latent utilities, of the joint posterior is as follows:
	\begin{equation}
		p(y,u,p|\Theta,\lambda,f) =\prod_{i\in\mathcal{N}}\prod_{t\in\mathcal{T}_i}p(y_{it}|u_{it})\left(\prod_{r\in\mathcal{R}}p(u_{irt}|\theta,\gamma,\lambda,f_i,p_{it})\prod_{j\in\mathcal{J}_p}p(p_{ijt}|\theta^p_j,\lambda,f_i)\right). \label{eq:full_likelihood}
	\end{equation}
	Details on the Bayesian probabilistic model are provided in Subsection \ref{sec:appendix_model}.
	
	\subsection{Priors}
	\label{sec:prior}
	
	Bayesian model specification is completed by assigning prior distributions to model parameters.
	For the equation parameters $\Theta=\left(\theta',\gamma',{\theta^p}'\right)'$, which consist of preference parameters $\theta=(\theta_1',\ldots,\theta_J')'$, bundle-effects parameters $\gamma=(\gamma_{(1,2)}',\ldots,\gamma_{(J-1,J)}')'$ and first-stage parameters $\theta^p=({\theta^p_1}',\ldots,{\theta^p_{J_p}}')'$, I apply a standard multivariate normal distribution priors to equation parameters, given by:
	\begin{equation*}
		\Theta\sim\mathcal{N}(m_\Theta,V_\Theta),
	\end{equation*}
	where $m_\Theta$ and $V_\Theta$ are hyperparameters.
	
	For the non-zero factor loadings in the factor loading matrix $\Lambda_t$, I assume {i.i.d.} normal distribution conjugate priors
	\begin{align*}
		\lambda_{jlt}&\sim\mathcal{N}(0,\sigma^2_\lambda), \text{for }j\in\mathcal{J}, l=1,\ldots,L, t\in\mathcal{T},\text{ and }\delta_{jlt}=1, \\
		\lambda^p_{jlt}&\sim\mathcal{N}(0,\sigma^2_\lambda), \text{for }j\in\mathcal{J}_p, l=1,\ldots,L, t\in\mathcal{T},\text{ and }\delta_{jlt}=1,
	\end{align*}
	with a common hyperparameter $\sigma^2_\lambda$.
	
	Note that the augmented latent utilities $u_{irt}$ do not require a prior, and the distribution of (augmented) latent factors $f_i\sim\mathcal{N}(\mathbf{0},\mathbf{I}_L)$ is specified as part of the model.
	
	\begin{table}[!h]
		\centering
		\caption{Guidelines for Choosing Number of Factors}
			\footnotesize
			\begin{tabular}{lcccccccccc}
\hline
      &       & RE   &       & \multicolumn{2}{c}{FA} &       & \multicolumn{4}{c}{TV-FA} \bigstrut\\
\cline{3-3}\cline{5-6}\cline{8-11}$J$     &       & \#    &       & $L$   & \#    &       & $L$   & \#    & $L+2$ & \# \bigstrut\\
\hline
\multicolumn{11}{l}{No Endogenous Regressor ($J_p=0$)} \bigstrut[t]\\
2     &       & 3     &       & 2     & 4     &       & 2     & $4T$  & 4     & $8T$ \\
3     &       & 6     &       & 2     & 6     &       & 2     & $6T$  & 4     & $12T$ \\
4     &       & 10    &       & 3     & 12    &       & 3     & $12T$ & 5     & $20T$ \\
5     &       & 15    &       & 3     & 15    &       & 3     & $15T$ & 5     & $25T$ \\
6     &       & 21    &       & 4     & 24    &       & 4     & $24T$ & 6     & $36T$ \\
\multicolumn{11}{l}{$J_p=J$ Endogenous Regressors} \\
2     &       & 10    &       & 3     & 12    &       & 3     & $12T$ & 5     & $20T$ \\
3     &       & 21    &       & 4     & 24    &       & 4     & $24T$ & 6     & $36T$ \\
4     &       & 36    &       & 5     & 40    &       & 5     & $40T$ & 7     & $56T$ \\
5     &       & 55    &       & 6     & 60    &       & 6     & $60T$ & 8     & $80T$ \\
6     &       & 78    &       & 7     & 84    &       & 7     & $84T$ & 9     & $108T$ \bigstrut[b]\\
\hline
\multicolumn{11}{p{0.5\textwidth}}{
Notes: This table provides guidelines for choosing the number of latent factors $L$ in Factor-Augmented (FA) and Time-Varying Factor-Augmented (TV-FA) models.
Columns labeled ``\#" indicate the number of free parameters modeling the error covariance. 
$J$ represents the number of goods, and $J_p$ denotes the number of endogenous regressors.
} \bigstrut[t]\\
\end{tabular}
		\label{tab:inference_L}
	\end{table}
	
	The model specification also involves choosing the number of factors $L$. Table \ref{tab:inference_L} provides guidelines for choosing the number of latent factors $L$ in Factor-Augmented (FA) and Time-Varying Factor-Augmented (TV-FA) models.
	As a general guideline, I match the number of free parameters modeling the error correlations $\Omega_{it}$ in the time-invariant factor-augmented model with that of the random-effects to ensure flexibility and that no restrictions are applied by the factor structure.
	When allowing the factor loading matrix $\Lambda_t$ to be time-varying, I add two more factors to the specification to capture the allowed time variations.

	\subsection{Posterior Inference}
	\label{sec:posterior}
	
	The vectorized model facilitates an efficient Markov-Chain Monte Carlo (MCMC) sampling scheme.
	The scheme for sampling from the posterior distribution involves the following steps:
	\begin{itemize}
		\item[1.] For $i\in\mathcal{N}$, $t\in\mathcal{T}_i$ and $r\in\mathcal{R}$, sample the latent utility from a truncated normal distribution
		\begin{equation*}
			u_{irt}|u_{i,-r,t},\Theta,\lambda,f_i\sim\begin{cases}
				\mathcal{N}(\mu(u_{irt}),1)\mathbf{1}(u_{irt}>\max(u_{i,-r,t})),\text{ if }y_{it}=r, \\
				\mathcal{N}(\mu(u_{irt}),1)\mathbf{1}(u_{irt}<\max(u_{i,-r,t})),\text{ if }y_{it}\neq r,
			\end{cases}
		\end{equation*}
		where $u_{i,-r,t}$ denotes the latent utilities in the choice set $\mathcal{R}$ other than $r$, and
		\begin{equation*}
			\mu(u_{irt})=\begin{cases}\sum_{j\in r}z_{ijt}'\theta_j+\sum_{j_1,j_2\in r}w_{i(j_1,j_2)t}'\gamma_{(j_1,j_2)}+\sum_{j\in r}\sum_{l=1}^L\lambda_{jlt}f_i, & \text{for }r\in\mathcal{R}\setminus0, \\
				0, & \text{for }r=0.
			\end{cases}
		\end{equation*}
		
		\item[2.] Sample the equation parameters $\Theta=(\theta',\gamma',{\theta^p}')'$ from
		\begin{equation*}
			\Theta|u,\lambda,f\sim\mathcal{N}(\bar{m}_\Theta,\bar{V}_\Theta),
		\end{equation*}
		where $\bar{V}_\Theta=(H_{zw}'H_{zw}+V_\Theta^{-1})^{-1}$, $\bar{m}_\Theta=\bar{V}_\Theta(H_{zw}'\tilde{y}^*+V_\Theta^{-1}m_\Theta)$ and $\tilde{y}^*=y^*-H_\Lambda f$.\footnote{See Appendix \ref{sec:appendix_mcmc} for an alternative sampler involves partially marginalizing over the factor $f$ in the spirit of \cite{wagner2023factor}.}
		
		\item[3.] Sample latent factors $f=(f_1',\ldots,f_N')'$ from
		\begin{equation*}
			f|u,\Theta,\lambda\sim\mathcal{N}(\bar{m}_f,\bar{V}_f),
		\end{equation*}
		where $\bar{V}_f=(H_\Lambda'H_\Lambda+\mathbf{I})^{-1}$, $\bar{m}_f=\bar{V}_f H_\Lambda'\tilde{\tilde{y}}^*$, and $\tilde{\tilde{y}}^*=y^*-H_{zw}\Theta$.
		
		\item[4.] Random sign switch of $f_l$ and $\lambda_l$ for $l=1,\ldots,L$ \citep{jacobi2016bayesian}.
		
		\item[5.] Boosting MCMC with marginal data augmentation \citep{van2001art}.
		
		\item[6.] Sample non-zero factor loadings $\lambda=\left(\{\{\{\lambda_{jlt}\}_{j\in\mathcal{J}|\delta_{jlt}=1},\{\lambda^p_{jlt}\}_{j\in\mathcal{J}_p|\delta_{jlt}=1}\}_{l=1}^L\}_{t\in\mathcal{T}}\right)'$ in $\Lambda$ from
		\begin{equation*}
			\lambda|u,\Theta,f\sim\mathcal{N}(\bar{m}_\lambda,\bar{V}_\lambda),
		\end{equation*}
		where $\bar{V}_\lambda=(H_f'H_f+\sigma^{-2}_\lambda\mathbf{I})^{-1}$, $\bar{m}_\lambda=\bar{V}_\lambda H_f'\tilde{\tilde{y}}^*$, and $\tilde{\tilde{y}}^*=y^*-H_{zw}\Theta$.
	\end{itemize}
	Step 1 is a standard Gibbs sampling step for the multinomial probit model (MNP), see \cite{mcculloch1994exact}.
	Since the error correlations are fully captured by the latent factors $f$ and factor loadings $\lambda$, the latent utilities $u_{irt}$ are conditionally independent.
	This simplifies the standard procedures.
	
	The model vectorization, introduced in Subsection \ref{sec:general_model}, enables the joint sampling of all equation parameters $\Theta=\left(\theta',\gamma',{\theta^p}'\right)'$ in Step 2.
	Similarly, it facilitates sampling all latent factors $f$ and factor loadings $\lambda$ in one block in Step 3 and 6, respectively.
	
	I develop the sampling Steps 3 to 6 for the factor model based on \cite{jacobi2016bayesian} and \cite{fruehwirthschnatter2023sparse}.
	Full details on this MCMC scheme are provided in Appendix \ref{sec:appendix_mcmc}.
	
	\subsection{Bayesian Predictive Approach for Price Elasticity}
	\label{sec:prediction}
	
	The key quantities of interest in bundle demand models are the price responses and complementarity, both of which are measured by the own- and cross-price elasticity of demand,
	\begin{equation*}
		\mathcal{E}_{jk}=\frac{\partial S_k}{S_k}/\frac{\partial p_j}{p_j},\qquad\text{for }j,k\in\mathcal{J},
	\end{equation*}
	and the counterfactual choice probability implied by policy changes,
	\begin{equation*}
		S_j^{cf}=\Pr(j\in y^{cf}),\qquad\text{for }j\in\mathcal{J}.
	\end{equation*}
	This subsection develops a Bayesian density predictions to estimates these quantities.
	
	The evaluation of both price elasticity and counterfactual choice probability hinges on the prediction of bundle choice $y_{it}$ given adjusted covariates $z_{ijt}^{cf}$ and $w_{i(j_1,j_2)t}^{cf}$.
	The full likelihood function \eqref{eq:full_likelihood} specifies how the bundle choice $y_{it}$ is generated:
	\begin{equation*}
		p(y_{it}^{cf}|u_{it}^{cf}) \prod_{r\in\mathcal{R}}p(u_{irt}^{cf}|\Theta,\Lambda_t,f_i,z_{it}^{cf},w_{it}^{cf}),
	\end{equation*}
	where $z_{it}=\{z_{ijt}^{cf}\}_{j\in\mathcal{J}}$ contains price and other observed covariates $z_{ijt}^{cf}=(p_{ijt}^{cf},{x_{ijt}^{cf}}')'$, and $w_{it}^{cf}=\{w_{i(j_1,j_2)t}^{cf}\}_{j_1,j_2\in\mathcal{J}}$.
	The densities $p(y_{it}|u_{it})$ and $p(u_{it}|\cdot)$ are defined by $y_{it}=\arg\max_{r\in\mathcal{R}}(u_{irt})$ and Equation \eqref{eq:general_utility_r_2}, respectively.
	
	Combine with the posterior density $p(\Theta,\Lambda,f|y)$, the predictive density for $y_{it}$ is given by
	\begin{equation*}
		p(y_{it}^{cf},u_{it}^{cf}|y,z_{it}^{cf},w_{it}^{cf})=\int p(y_{it}^{cf}|u_{it}^{cf}) \prod_{r\in\mathcal{R}} p(u_{irt}^{cf}|\Theta,\Lambda_t,f_i,z_{it}^{cf},w_{it}^{cf}) p(\Theta,\Lambda_t,f_i|y)d (\Theta,\Lambda,f).
	\end{equation*}
	
	The choice probabilities of consuming good $j$ and bundle $r$ are formally defined as
	\begin{equation*}
		S_{j}^{cf}=\int\int\Pr(j\in y_{it}^{cf})p(i)p(t)didt,\ \text{and }S_{r}^{cf}=\int\int\Pr(y_{it}^{cf}=r)p(i)p(t)didt,
	\end{equation*}
	where the integrals are over the empirical distribution of the individuals $i\in\mathcal{N}$ and time periods $t\in\mathcal{T}$.
	It specifically calculated by
	\begin{equation*}
		S_{j}^{cf}=\frac{1}{\sum_{i}^N T_i}\sum_{i=1}^N\sum_{t=1}^{T_i}\mathbf{1}(j\in y_{it}^{cf}),\ \text{and }S_{r}^{cf}=\frac{1}{\sum_{i}^N T_i}\sum_{i=1}^N\sum_{t=1}^{T_i}\mathbf{1}(y_{it}^{cf}=r),
	\end{equation*}
	as functions of the predicted bundle choice $y_{it}$.
	Their predictive densities are readily computed.
	
	The evaluation of bundle and good price elasticity uses a two-sided numerical differentiation approach based on the above Bayesian predictive approach. Specifically, I simulate the bundle and good choice probability at three points, (1) decrease the observed price $p_{ijt}$ by 5\%, $p_{ijt}^\text{backward}=p_{ijt}\times(1-0.05)$, (2) original price $p_{ijt}$, and (3) increase by 5\%, $p_{ijt}^\text{forward}=p_{ijt}\times(1+0.05)$. Then I calculate the elasticities by
	\begin{align*}
		\mathcal{E}_{jr}&=\frac{\partial S_r}{S_r}\frac{p_j}{\partial p_j}=\frac{S_{r}^{cf}(\{\{p_{ijt}^\text{forward}\}_{t\in\mathcal{T}_i}\}_{i\in\mathcal{N}})-S_{r}^{cf}(\{\{p_{ijt}^\text{backward}\}_{t\in\mathcal{T}_i}\}_{i\in\mathcal{N}})}{S_{r}^{cf}(\{\{p_{ijt}\}_{t\in\mathcal{T}_i}\}_{i\in\mathcal{N}})}\frac{1}{10\%}, \\
		\mathcal{E}_{jk}&=\frac{\partial S_k}{S_k}\frac{p_j}{\partial p_j}=\frac{S_{k}^{cf}(\{\{p_{ijt}^\text{forward}\}_{t\in\mathcal{T}_i}\}_{i\in\mathcal{N}})-S_{k}^{cf}(\{\{p_{ijt}^\text{backward}\}_{t\in\mathcal{T}_i}\}_{i\in\mathcal{N}})}{S_{k}^{cf}(\{\{p_{ijt}\}_{t\in\mathcal{T}_i}\}_{i\in\mathcal{N}})}\frac{1}{10\%},
	\end{align*}
	where bundle elasticity $\mathcal{E}_{jr}$ is the percentage change in the probability use of bundle $r$ with respect to 1\% increase in the price of good $j$. Similarly, substance elasticity $\mathcal{E}_{jk}$ is the percentage change in the probability use of good $k$ with respect to 1\% increase in the price of good $j$. For both equations, the right hand side is the multiplication of the differentiation respect to a 10\% change (backward plus forward) in price and normalization of the 10\% change.
	
	\section{Simulation Studies}
	\label{sec:simulation}
	
	I explore the performance of the proposed model, the time-varying factor-augmented bundle demand model with endogenous regressors, comparing it to the benchmark time-invariant random-effects and factor-augmented models with or without endogenous regressors. Furthermore, I explore the asymptotic behavior of the proposed model with various sample sizes.
	
	\subsection{Simulation Setup}
	\label{sec:simulation_setup}
	
	For each data set, there are $J=3$ goods and $J_p=3$ good prices $p_{ijt}$ are endogenous. The covariates (and parameters) are simulated (and set to) mimicking the data application.
	
	For the utility equations, the price coefficient is set to $\alpha=-1$.
	It is common to all three good utilities.
	The remaining preference parameters are $\beta_1=(1,0.2,0.1)'$, $\beta_2=(2,0.2,0.1)'$ and $\beta_3=(2,0.1,0.05)'$ for three goods, respectively.
	Similarly, there is a common parameter for all bundle effects, $\tilde{\gamma}=0.05$.
	The remaining bundle-effects parameters are bundle-specific intercepts, $\tilde{\gamma}_{(1,2)}=2$, $\tilde{\gamma}_{(1,3)}=0$ and $\tilde{\gamma}_{(2,3)}=-1$.
	Intuitively, good 1 and 2 are complements, 1 and 3 are independent, and 2 and 3 are substitutes, according to the extra utility or disutility from the bundle effects.
	For the first-stage equations, the parameters are set to $\theta^p_1=(0,1,0.5,0,0.01)'$, $\theta^p_2=(0,1,0.5,0,0)'$ and $\theta^p_3=(0,1,0.5,0,-0.01)'$ for $j=1,2,3$, respectively.
	
	The error correlations are induced by two latent factors $f_i\sim\mathcal{N}(\mathbf{0},\mathbf{I}_2)$ and time-varying factor loadings $\Lambda_t=(\lambda_{1t},\lambda_{2t})$ with $\lambda_{1t}=(1,0,-1,1,0,-1)'$ and $\lambda_{2t}\sim\mathcal{N}(\mathbf{0},\mathbf{I}_6)$.
	The six rows of $\Lambda_t$ correspond to the three good utility and three first-stage equations, respectively.
	Note that the $\Lambda_t$ cannot be recovered due to rotational invariance.
	It is a means of introducing error correlations, with the same spirit of the model where the loading is a means of capturing error correlations.
	
	The observed covariates in the utility equations are $z_{ijt}=(p_{ijt},1,x_{1i},x_{2i})'$, for $j\in\mathcal{J}$, where $p_{ijt}$ is the (endogenous) price, $1$ corresponds to the intercept, $x_{1i}\sim\mathcal{N}(10,0.7225)$, and $x_{2i}\sim\mathcal{TN}_{(0,\infty)}(10,36)$.
	The bundle-effects covariates are $w_{i(j_1,j_2)t}=(\tilde{w}_i,1)'$ for $j_1,j_2\in\mathcal{J}$, interacting with $\gamma=(\tilde{\gamma},\tilde{\gamma}_{(j_1,j_2)})'$, where $\tilde{w}_i=1,2,\ldots,6$ mimicking the family size.
	The instruments and exogenous regressors are $z^p_{ijt}=(p_{jt},1,z_{ij},x_{1i},x_{2i})'$ where $p_{jt}$ mimics the market-level price $p_{1t}\sim\mathcal{N}(7,0.04)$, $p_{2t}\sim\mathcal{N}(6,0.01)$ and $p_{3t}\sim\mathcal{N}(5,0.01)$.
	The instruments are $z_{ijt}\sim\mathcal{N}(0,1)$, for individual $i$, time $t$, and good $j\in\mathcal{J}_p$.

	\subsection{Comparison Across Models} 
	\label{sec:simulation_model}
	
	This subsection compare the proposed time-varying factor-augmented bundle demand model with endogenous regressors with the benchmark time-invariant random-effects and factor-augmented models with or without endogenous regressors.
	The true data-generating process (DGP) is outlined in Subsection \ref{sec:simulation_setup}.
	The $J_p=3$ prices are endogenous and the unobserved tastes and the unobserved utility-price confounders are time-varying.
	So the time-varying factor-augmented (TV-FA) model corresponds to the true DGP.
	I simulate 50 Monte Carlo trials.
	For each trial, the sample contains $N=1000$ individuals and $T=12$ time periods, which is close to the dimension of the real data application.
	For each sample and model, the MCMC simulates 10,000 draws for use after a 10,000-draw burn-in period.
	
	The key quantities of interest in bundle demand models are price responses and complementarity, both measured by own- and cross-price elasticity of demand.
	Therefore, the comparison focuses on the estimated price elasticity, specifically the distance between the true elasticities evaluated with the true model parameters and the estimated elasticities evaluated with the estimated parameters.
	
	Table \ref{tab:elasticity_model} reports the true price elasticity and the differences between the true and estimated elasticities from the five models.
	The first column shows the true values of elasticities from the true DGP, bundle demand model with endogenous regressors (Endo) and time-varying (TV) unobserved tastes and the unobserved utility-price confounders.
	The remaining five columns correspond to the time-invariant (TI) random-effects (RE) exogenous (Exo) model, TI factor-augmented (FA) exogenous model, TI-FA model with endogenous regressors (Endo), time-varying (TV) FA exogenous (Exo) model, and lastly the correctly specified time-varying factor-augmented (TV-FA) model with endogenous regressors (Endo).
	The root mean squared errors (RMSE) are calculated from the squared difference between posterior mean and true value of price elasticity for each Monte Carlo trail.
	
	This simulation study suggests that the correctly specified and most flexible TV-FA Endo model outperforms the four benchmark models.
	The comparison between the exogenous TV-FA Exo and endogenous TV-FA Endo models highlights the importance of controling for price endogeneity when estimating complementarity.
	The comparison between the time-invariant FA Endo (TI) and the time-varying FA Endo models highlights that accounting for time-varying unobserved tastes and the unobserved utility-price confounders is also important in the bundle demand models.
	
	\begin{table}[!h]
		\centering
		\caption{True Price Elasticities and Root Mean Squared Errors (RMSE) of Estimated Elasticities by Models Compared to True Values}
		\resizebox{\textwidth}{!}{
			\footnotesize
			\begin{tabular}{lcccccccc}
\hline
      &       & True  &       & \multicolumn{5}{c}{Root Mean Squared Error (RMSE)} \bigstrut\\
\cline{5-9}      &       & Elasticities &       & RE Exo (TI) & FA Exo (TI) & FA Endo (TI) & TV-FA Exo & \textbf{TV-FA Endo} \bigstrut\\
\hline
\multicolumn{9}{l}{Own-Price Elasticity} \bigstrut[t]\\
$\mathcal{E}_{11}$ &       & -4.020 &       & 0.582 & 0.674 & 0.383 & 0.582 & 0.143 \\
$\mathcal{E}_{22}$ &       & -2.660 &       & 0.337 & 0.369 & 0.259 & 0.184 & 0.078 \\
$\mathcal{E}_{33}$ &       & -2.688 &       & 0.403 & 0.506 & 0.306 & 0.375 & 0.111 \\
\multicolumn{9}{l}{Cross-Price Elasticity w.r.t. Price 1} \\
$\mathcal{E}_{12}$ &       & -0.718 &       & 0.103 & 0.126 & 0.106 & 0.109 & 0.045 \\
$\mathcal{E}_{13}$ &       & 0.219 &       & 0.170 & 0.147 & 0.150 & 0.170 & 0.056 \\
\multicolumn{9}{l}{Cross-Price Elasticity w.r.t. Price 2} \\
$\mathcal{E}_{21}$ &       & -1.350 &       & 0.212 & 0.269 & 0.195 & 0.196 & 0.080 \\
$\mathcal{E}_{23}$ &       & 0.662 &       & 0.170 & 0.173 & 0.159 & 0.120 & 0.069 \\
\multicolumn{9}{l}{Cross-Price Elasticity w.r.t. Price 3} \\
$\mathcal{E}_{31}$ &       & 0.176 &       & 0.135 & 0.119 & 0.120 & 0.133 & 0.059 \\
$\mathcal{E}_{32}$ &       & 0.281 &       & 0.072 & 0.073 & 0.072 & 0.058 & 0.036 \bigstrut[b]\\
\hline
\multicolumn{9}{p{\textwidth}}{Notes:
This table reports the true price elasticity and the differences between the true and estimated elasticities from the five models.
The root mean squared errors (RMSE) represent the differences between the true value and the posterior mean (obtained from 10,000 MCMC draws after a 10,000 draw burn-in period) of the evaluated price elasticity, based on 50 Monte Carlo trials.
For each Monte Carlo trial, I simulate $N=1000$ individuals and $T=12$ time periods.
The \textbf{TV-FA Endo} model is the most flexible model proposed in this paper and corresponds to the true data-generating process outlined in Subsection \ref{sec:simulation_setup}.
``RE" stands for random-effects, ``FA" for factor-augmented, ``TI" for time-invariant (applicable to both random effects or factor loadings), ``TV" for time-varying factor loadings, ``Exo" for the exogenous model without endogenous regressors, and ``Endo" for the endogenous model with $J_p=J=3$ endogenous prices.
Price elasticity $\mathcal{E}_{jk}$ is the percentage change in the choice probability of using good $k$ in response to a percentage increase in the price of good $j$.
} \bigstrut[t]\\
\end{tabular}}
		\label{tab:elasticity_model}
	\end{table}
	
	\subsection{Asymptotic Results}
	\label{sec:simulation_sample_size}
	
	This subsection investigate the asymptotic behavior of the proposed model.
	The true data-generating process (DGP) is outlined in Subsection \ref{sec:simulation_setup}.
	The $J_p=3$ prices are endogenous and the unobserved tastes and the unobserved utility-price confounders are time-varying.
	I consider five sample sizes with $N=100,1000,1000$ individuals and $T=6,12,24$ time periods.
	For the sample size, 50 Monte Carlo trials are simulated and estimated by the correctly specified time-varying factor-augmented (TV-FA) endogenous model.
	
	Table \ref{tab:elasticity_sample_size} reports the asymptotic results.
	The first column shows the true values of elasticities from the true DGP, bundle demand model with endogenous regressors (Endo) and time-varying (TV) unobserved tastes and the unobserved utility-price confounders.
	The remaining five columns correspond to five sample sizes.
	The results show clear pictures of convergence.
	As number of individuals $N$ or number of time periods $T$ increases, the distances between the true price elasticities and the estimated elasticities shrink.
	
	\begin{table}[!h]
		\centering
		\caption{True Price Elasticities and Root Mean Squared Errors (RMSE) of Estimated Elasticities by Sample Sizes Compared to True Values}
		\resizebox{0.8\textwidth}{!}{
			\footnotesize
			\begin{tabular}{lcccccccc}
\hline
      &       &       &       & \multicolumn{5}{c}{Root Mean Squared Error (RMSE)} \bigstrut\\
\cline{5-9}      &       & True  &       & $N=100$ & $N=1000$ & $N=10000$ & $N=1000$ & $N=1000$ \bigstrut[t]\\
      &       & Elasticities &       & $T=12$ & $T=12$ & $T=12$ & $T=6$ & $T=24$ \bigstrut[b]\\
\hline
\multicolumn{9}{l}{Own-Price Elasticity} \bigstrut[t]\\
$\mathcal{E}_{11}$ &       & -4.020 &       & 0.479 & 0.143 & 0.047 & 0.240 & 0.096 \\
$\mathcal{E}_{22}$ &       & -2.660 &       & 0.277 & 0.078 & 0.026 & 0.117 & 0.052 \\
$\mathcal{E}_{33}$ &       & -2.688 &       & 0.350 & 0.111 & 0.033 & 0.139 & 0.066 \\
\multicolumn{9}{l}{Cross-Price Elasticity w.r.t. Price 1} \\
$\mathcal{E}_{12}$ &       & -0.718 &       & 0.152 & 0.045 & 0.016 & 0.067 & 0.026 \\
$\mathcal{E}_{13}$ &       & 0.219 &       & 0.162 & 0.056 & 0.017 & 0.068 & 0.027 \\
\multicolumn{9}{l}{Cross-Price Elasticity w.r.t. Price 2} \\
$\mathcal{E}_{21}$ &       & -1.350 &       & 0.288 & 0.080 & 0.033 & 0.134 & 0.054 \\
$\mathcal{E}_{23}$ &       & 0.662 &       & 0.220 & 0.069 & 0.023 & 0.107 & 0.043 \\
\multicolumn{9}{l}{Cross-Price Elasticity w.r.t. Price 3} \\
$\mathcal{E}_{31}$ &       & 0.176 &       & 0.169 & 0.059 & 0.014 & 0.077 & 0.028 \\
$\mathcal{E}_{32}$ &       & 0.281 &       & 0.125 & 0.036 & 0.012 & 0.050 & 0.023 \bigstrut[b]\\
\hline
\multicolumn{9}{p{0.8\textwidth}}{Notes:
This table reports the true price elasticity and the differences between the true and estimated elasticities from the correctly specified time-varying factor-augmented (TV-FA) bundle demand model.
The five columns correspond to five sample size setting with number of individuals $N=100,1000,1000$ and time periods $T=6,12,24$.
The root mean squared errors (RMSE) represent the differences between the true value and the posterior mean (obtained from 10,000 MCMC draws after a 10,000 draw burn-in period) of the evaluated price elasticity, based on 50 Monte Carlo trials.
Price elasticity $\mathcal{E}_{jk}$ is the percentage change in the choice probability of using good $k$ in response to a percentage increase in the price of good $j$.
} \bigstrut[t]\\
\end{tabular}}
		\label{tab:elasticity_sample_size}
	\end{table}

	\section{Sugary Drink Tax under Complementarities}
	\label{sec:soda_application}
	
	This section applies the proposed model to examine the implications of implementing a sugary drink tax (also known as a soda tax) within the context of complementarities.
	
	\subsection{Data and Model Specification}
	\label{sec:data_and_specification}
	
	I combine household-level and store-level information on carbonated beverages, milk, and salty snacks from the IRI Marketing Data Set \citep{bronnenberg2008database}.
	The household-level data, collected from two BehaviorScan markets in Pittsfield, Massachusetts, and Eau Claire, Wisconsin, include records of shopping trips, purchases, and demographic information.
	Store-level data encompass transactions for each product, identified by Universal Product Code (UPC), across each store and week.
	Additionally, this data set provides product attributes across five levels (category, small category, parent company, vendor, brand) and the equivalent volume of the package.
	
	\begin{table}[htb]
		\centering
		\caption{Goods, Choice Probabilities and Prices}
		\resizebox{\textwidth}{!}{
			\footnotesize
\begin{tabular}{lllccc}
\hline
      &       &       & \multicolumn{2}{c}{Proportion of Purchase} & Price in US\$ \bigstrut\\
\cline{4-5}$j$   & Good  & Subcategory & Weekly & Ever & (per 6 servings) \bigstrut\\
\hline
1     & Sugary Soft Drinks & Regular Soft Drinks & 18.79\% & 84.60\% & 2.06 \bigstrut[t]\\
2     & Diet Soft Drinks & Low Calorie Soft Drinks & 18.88\% & 75.31\% & 2.05 \\
3     & Carbonated Water & Seltzer, Tonic Water, Club Soda & 3.43\% & 24.69\% & 2.39 \\
4     & Milk Drinks & Flavored Milk, Eggnog, Milkshakes, … & 10.19\% & 65.48\% & 3.40 \\
5     & Salty Snacks & Cheese Snacks, Potato Chips, … & 35.48\% & 97.01\% & 1.75 \bigstrut[b]\\
\hline
\multicolumn{6}{p{1.08\textwidth}}{Notes: This table defines five goods by product subcategory (Column 1-3), and presents weekly purchase probabilities based on 163,434 choices (household-week combinations) by 4,143 households over 52 weeks in 2011 (Column 4), the proportion of households that purchased each good at least once during the year (Column 5), and the average prices in U.S. dollars per 72 fluid ounces for $j=1,\ldots,4$ and per 6 ounces for $j=5$ (Column 6).} \bigstrut[t]\\
\end{tabular}}
		\label{tab:soda_choice_prob_goods}
	\end{table}
	
	Table \ref{tab:soda_choice_prob_goods} defines the $J=5$ goods.
	The first good, identified as a collection of sugary soft drinks, is targeted by the soda tax.
	This category encompasses beverages sweetened with cane, corn, or other sugars.
	The second good includes diet soft drinks, which are low-calorie beverages that are either sugar-free or sweetened with alternatives such as Aspartame, Sucralose, or Stevia.
	Regions such as France, the Philippines, the United Arab Emirates, and Philadelphia (Pennsylvania, United States) impose the tax on artificially sweetened beverages at the same rate as sugar-sweetened beverages.
	Conversely, these beverages are excluded from the tax in other areas, such as Canada and Berkeley (California, United States).
	
	The third good, carbonated water, comprises seltzer, tonic water, and club soda, which are typically exempt from the tax.
	The fourth good includes milk drinks, such as flavored milk, eggnog, buttermilk, milkshakes, and non-dairy drinks, also generally not subject to the tax.
	These are included in the analysis as alternative sugar sources.
	The fifth good, consisting of salty snacks (e.g., cheese snacks, corn snacks, potato chips, ready-to-eat popcorn, tortilla chips), is considered a potential complement to soft drinks.
	This section explores the substitution and complementarity among these five goods.
	
	The sample consists of 4,143 households that purchased any of the five goods at least once over the 52 weeks of 2011.
	Observed choices are defined as combinations of household and week.
	A household is considered to have made a choice decision if it visited a store and purchased any products, including but not limited to the five goods, within a given week.
	This results in a total of 163,434 choice decisions.
	The fourth column of Table \ref{tab:soda_choice_prob_goods} presents the probability of purchasing each of the five goods in any week,
	while the fifth column shows the proportion of the 4,143 households that purchased each good at least once throughout the year.
	
	Table \ref{tab:soda_choice_prob_bundles} defines all 32 possible bundles of goods, ranging from the empty bundle (no purchase) to combinations of one to five goods.
	The last column reports the weekly choice probability for each bundle based on 163,434 choices.
	Among these choices, 43.78\% involve no purchase, 32.96\% chose a single good, 16.82\% chose two goods, 5.61\% chose three goods, 0.81\% chose four goods, 0.02\% chose all goods.
	
	\begin{table}[htb]
		\centering
		\caption{Bundle Choice Probabilities}
		\resizebox{\textwidth}{!}{
			\footnotesize
\begin{tabular}{llcccccc}
\hline
      & $r$   & \multicolumn{5}{c}{Bundle / Choice}   & Probability \bigstrut[b]\\
\hline
\multicolumn{8}{l}{\textbf{Empty Bundle}} \bigstrut[t]\\
      & $0\text{ or }\varnothing$ & \multicolumn{5}{c}{Empty Bundle (Outside Option)} & 43.78\% \\
\multicolumn{8}{l}{\textbf{One-Good Bundles}} \\
      & $1$ & Sugary Soft Drinks &       &       &       &       & 5.41\% \\
      & $2$ &       & Diet Soft Drinks &       &       &       & 5.74\% \\
      & $3$ &       &       & Carbonated Water &       &       & 1.25\% \\
      & $4$ &       &       &       & Milk Drinks &       & 3.49\% \\
      & $5$ &       &       &       &       & Salty Snacks & 17.07\% \\
\multicolumn{8}{l}{\textbf{Two-Good Bundles}} \\
      & $(1,2)$ & Sugary Soft Drinks & Diet Soft Drinks &       &       &       & 2.26\% \\
      & $(1,3)$ & Sugary Soft Drinks &       & Carbonated Water &       &       & 0.22\% \\
      & $(1,4)$ & Sugary Soft Drinks &       &       & Milk Drinks &       & 0.76\% \\
      & $(1,5)$ & Sugary Soft Drinks &       &       &       & Salty Snacks & 4.83\% \\
      & $(2,3)$ &       & Diet Soft Drinks & Carbonated Water &       &       & 0.19\% \\
      & $(2,4)$ &       & Diet Soft Drinks &       & Milk Drinks &       & 0.75\% \\
      & $(2,5)$ &       & Diet Soft Drinks &       &       & Salty Snacks & 4.76\% \\
      & $(3,4)$ &       &       & Carbonated Water & Milk Drinks &       & 0.08\% \\
      & $(3,5)$ &       &       & Carbonated Water &       & Salty Snacks & 0.79\% \\
      & $(4,5)$ &       &       &       & Milk Drinks & Salty Snacks & 2.17\% \\
\multicolumn{8}{l}{\textbf{Three-Good Bundles}} \\
      & $(1,2,3)$ & Sugary Soft Drinks & Diet Soft Drinks & Carbonated Water &       &       & 0.11\% \\
      & $(1,2,4)$ & Sugary Soft Drinks & Diet Soft Drinks &       & Milk Drinks &       & 0.43\% \\
      & $(1,2,5)$ & Sugary Soft Drinks & Diet Soft Drinks &       &       & Salty Snacks & 2.79\% \\
      & $(1,3,4)$ & Sugary Soft Drinks &       & Carbonated Water & Milk Drinks &       & 0.02\% \\
      & $(1,3,5)$ & Sugary Soft Drinks &       & Carbonated Water &       & Salty Snacks & 0.26\% \\
      & $(1,4,5)$ & Sugary Soft Drinks &       &       & Milk Drinks & Salty Snacks & 0.89\% \\
      & $(2,3,4)$ &       & Diet Soft Drinks & Carbonated Water & Milk Drinks &       & 0.02\% \\
      & $(2,3,5)$ &       & Diet Soft Drinks & Carbonated Water &       & Salty Snacks & 0.23\% \\
      & $(2,4,5)$ &       & Diet Soft Drinks &       & Milk Drinks & Salty Snacks & 0.80\% \\
      & $(3,4,5)$ &       &       & Carbonated Water & Milk Drinks & Salty Snacks & 0.06\% \\
\multicolumn{8}{l}{\textbf{Four-Good Bundles}} \\
      & $(1,2,3,4)$ & Sugary Soft Drinks & Diet Soft Drinks & Carbonated Water & Milk Drinks &       & 0.02\% \\
      & $(1,2,3,5)$ & Sugary Soft Drinks & Diet Soft Drinks & Carbonated Water &       & Salty Snacks & 0.13\% \\
      & $(1,2,4,5)$ & Sugary Soft Drinks & Diet Soft Drinks &       & Milk Drinks & Salty Snacks & 0.62\% \\
      & $(1,3,4,5)$ & Sugary Soft Drinks &       & Carbonated Water & Milk Drinks & Salty Snacks & 0.03\% \\
      & $(2,3,4,5)$ &       & Diet Soft Drinks & Carbonated Water & Milk Drinks & Salty Snacks & 0.02\% \\
\multicolumn{8}{l}{\textbf{Full Bundle}} \\
      & $(1,2,3,4,5)$ & Sugary Soft Drinks & Diet Soft Drinks & Carbonated Water & Milk Drinks & Salty Snacks & 0.02\% \bigstrut[b]\\
\hline
\multicolumn{8}{p{1.15\textwidth}}{Notes: This table defines 32 possible bundles of goods, ranging from the empty bundle (no purchase) to combinations of one to five goods. The last column reports the choice probability for each bundle, based on 163,434 choices (household-week combinations) made by 4,143 households over 52 weeks in 2011.} \bigstrut[t]\\
\end{tabular}
}
		\label{tab:soda_choice_prob_bundles}
	\end{table}
	
	Table \ref{tab:soda_choice_prob_goods} includes the average prices of the five goods in the last column.
	I construct $\textit{price}_{ijt}$ for good $j$ faced by household $i$ in week $t$, using $i$'s complete purchase history, $i$'s shopping trips in week $t$, and the average product prices at the stores visited by $i$ during week $t$.
	The steps for constructing $\textit{price}{ijt}$ are as follows:
	
	First, an aggregate product basket for each good is defined, based on 5-level product categories and the package's equivalent volume.
	The aggregate baskets for the five goods consist of 292, 228, 32, 50, and 642 products, respectively.
	These numbers are fewer than the total number of unique products identified by UPC codes, as a product can have multiple UPCs across different stores and cities.
	Table \ref{tab:soda_price_construction} outlines the definitions of the aggregate product baskets for the five goods.
	Each household $i$'s specific product basket for good $j$ is a subset of this aggregate basket.
	The purchase record of household $i$ in 2011 then determines the weight of each product in their basket, proportional to the total volume of good $j$ that the household purchased, relative to all volumes purchased of that good."
	
	Second, for each shopping trip of household $i$ in week $t$, the effective price of good $j$ is equal to the weighted average prices of the products in $i$'s basket, with weights derived from $i$'s basket.
	If household $i$ makes multiple shopping trips in week $t$, then $\textit{price}_{ijt}$ equals the lowest of these weighted average prices across all trips.
	
	Prices $\text{price}_{ijt}$ might be endogenous due to three factors.
	First, the product baskets chosen by household $i$ are likely correlated with $i$'s preferences through unobserved characteristics.
	Second, the shopping trips of household $i$ may also be linked to their preferences.
	Lastly, demand shocks can influence households' preferences for products as well as the prices of these products.
	To address endogeneity, I consider two sets of instrumental variables.
	
	The first set includes Hausman-type instruments, following the approach of \citep{nevo2000practitioner}.
	For Pittsfield, the instruments comprise the unit prices of the same aggregate product basket in three surrounding cities in the market-level data: Boston, Hartford, and New York.
	Similarly, the instruments for Eau Claire are derived from surrounding cities, including Chicago, Green Bay, Indianapolis, Milwaukee, Minneapolis/St. Paul, and Omaha.
	These instrumental variables are correlated with the product prices but are unlikely to be influenced by demand shocks occurring in Eau Claire or Pittsfield, given the geographical and market separation.
	
	Furthermore, another instrument is the number of different stores that household $i$ visited in week $t$.
	Visiting more stores is assumed to expose the household to lower effective prices.
	Meanwhile, the number of different stores visited, distinct from the number of shopping trips, is presumed to be uncorrelated with the household's preferences.
	
	I specify the good utility as
	\begin{multline*}
		\bar{u}_{ijt}=\alpha_j\times\textit{price}_{ijt} + \beta_{j0} + \beta_{j1}\times\textit{\# of trips}_{it} + \beta_{j2}\times\textit{income}_i + \beta_{j3}\times\textit{family size}_i \\ + \beta_{j4}\times\textit{child 0-5}_i + \beta_{j5}\times\textit{child 6-11}_i + \beta_{j6}\times\textit{child 12-17}_i + \beta_{j7}\times\textit{Pittsfield}_i + \nu_{ijt} , \\ \qquad\text{for }j=1,2,\ldots,5,
	\end{multline*}
	where $\textit{price}_{ijt}$ represents the price of good $j$ faced by individual $i$ in week $t$.
	As the goods differ in nature and thus exhibit varying price responses, I allow the price coefficient $\alpha_j$ to vary across goods.
	$\textit{\# of trips}_{it}$ denotes the number of shopping trips by household $i$ in week $t$.
	A visit is recorded if the household makes any purchase, regardless of whether it includes any of the goods of interest.
	$\textit{income}_i$ is the self-reported annual household income, expressed in log thousands of U.S. dollars.
	$\textit{family size}_i$ represents the family size, which ranges from 1 to 8.
	$\textit{child 0-5}_i$, $\textit{child 6-11}_i$, and $\textit{child 12-17}_i$ are dummy variables indicating the presence of children within those age ranges in the household.
	It is noted that more than one of these dummy variables can be equal to one, indicating the presence of multiple children in different age categories.
	$\textit{Pittsfield}_i$ is a dummy variable identifying households from Pittsfield; otherwise, the household resides in Eau Claire.
	Additional household characteristics, such as the age and gender of the household head, level of education, and employment status, were considered but did not yield statistically significant coefficients or alter the outcomes.
	
	The bundle effects are specified as follows:
	\begin{align*}
		\Gamma_{irt}&=\sum_{j_1,j_2\in r}\Gamma_{i(j_1,j_2)t}, \\
		\Gamma_{i(j_1,j_2)t}&=\gamma_{(j_1,j_2)0} + \gamma_{(j_1,j_2)1}\times\textit{income}_i + \gamma_{(j_1,j_2)2}\times\textit{child 0-5}_i \\ &\quad + \gamma_{(j_1,j_2)3}\times\textit{child 6-11}_i + \gamma_{(j_1,j_2)4}\times\textit{child 12-17}_i \\ &\quad + \gamma_{(j_1,j_2)5}\times\textit{family size}_i, \qquad\text{for }j_1,j_2=1,\ldots,5\text{ and }j_1\neq j_2,
	\end{align*}
	where $\gamma_{(j_1,j_2)0}$ captures the baseline additional utility or disutility from the joint consumption of goods $j_1$ and $j_2$.
	The other coefficients $\gamma_{(j_1,j_2)k}$ account for the heterogeneities in preferences.

	\subsection{Estimates of Model Parameters}
	
	\begin{table}[htb]
		\centering
		\caption{Estimates of Good Utility Parameters}
		\resizebox{\textwidth}{!}{
			\footnotesize
\begin{tabular}{llccccc}
\hline
      & $j = $ & 1     & 2     & 3     & 4     & 5 \bigstrut[t]\\
      &       & Sugary Soft Drinks & Diet Soft Drinks & Carbonated Water & Milk Drinks & Salty Snacks \bigstrut[b]\\
\hline
\multicolumn{7}{l}{\boldmath{}\textbf{Good Utility $\bar{u}_{ijt}$}\unboldmath{}} \bigstrut[t]\\
      & Price & -0.330*** & -0.422*** & -0.204*** & -0.356*** & -1.193*** \\
      &       & (0.016) & (0.019) & (0.043) & (0.016) & (0.023) \\
      &       &       &       &       &       &  \\
      & Constant & -0.888*** & -2.656*** & -6.464*** & -1.773*** & -0.691*** \\
      &       & (0.204) & (0.217) & (0.253) & (0.199) & (0.171) \\
      & \# of Trips & 0.128*** & 0.121*** & 0.096*** & 0.137*** & 0.152*** \\
      &       & (0.004) & (0.004) & (0.008) & (0.005) & (0.003) \\
      & Income & -0.062*** & 0.119*** & 0.253*** & 0.020 & 0.141*** \\
      &       & (0.017) & (0.020) & (0.022) & (0.018) & (0.017) \\
      & Family Size & 0.123*** & 0.080*** & 0.012 & 0.096*** & 0.086*** \\
      &       & (0.018) & (0.024) & (0.031) & (0.018) & (0.013) \\
      &       &       &       &       &       &  \\
      & Child 0-5 & 0.126 & -0.129 & -0.033 & 0.097 & 0.025 \\
      &       & (0.093) & (0.131) & (0.139) & (0.092) & (0.075) \\
      & Child 6-11 & 0.093 & -0.014 & -0.189 & 0.217*** & 0.149*** \\
      &       & (0.059) & (0.098) & (0.110) & (0.062) & (0.052) \\
      & Child 12-17 & 0.187*** & -0.033 & -0.192 & 0.234*** & 0.126*** \\
      &       & (0.059) & (0.065) & (0.155) & (0.061) & (0.044) \\
      &       &       &       &       &       &  \\
      & Pittsfield & -0.026 & -0.212*** & 0.883*** & -0.220*** & 0.107** \\
      &       & (0.055) & (0.065) & (0.076) & (0.041) & (0.049) \bigstrut[b]\\
\hline
\multicolumn{7}{p{1.05\textwidth}}{Notes:
This table presents the posterior means and standard deviations (in parentheses) of the good utility parameters from the main specification by the main model, the time-varying factor-augmented (TV-FA) bundle demand model with endogenous regressors.
The five columns correspond to the five goods.
*** $p<0.01$, ** $p<0.05$, * $p<0.1$.
The significance levels are based on the posterior credibility intervals.
} \bigstrut[t]\\
\end{tabular}}
		\label{tab:soda_estimates_goods}
	\end{table}
	
	Table \ref{tab:soda_estimates_goods} presents the posterior means and standard deviations (in parentheses) of the utility parameters for various goods, as derived from the time-varying factor-augmented (TV-FA) bundle demand model with endogenous regressors.
	The results are in line with expectations, with price coefficients being negative and precisely estimated, evidenced by small posterior standard deviations.
	Additionally, an observation is the impact of household income on consumption preferences.
	Wealthier households demonstrate a shift away from sugary soft drinks towards diet soft drinks and carbonated water, as shown by the negative income effects for sugary soft drinks and positive effects for diet soft drinks and carbonated water.
	Furthermore, households with children aged 0-5 do not exhibit significant differences in consumption preferences among the five goods.
	However, the presence of older children (ages 6-11 and 12-17) markedly increases the utility derived from milk drinks and salty snacks.
	
	Table \ref{tab:soda_estimates_bundles} presents the posterior means and standard deviations (in parentheses) of the bundle-effects parameters, from the time-varying factor-augmented (TV-FA) bundle demand model with endogenous regressors.
	Each column in the top, middle, and bottom panels corresponds to a pair of goods, $j_1$ and $j_2$, illustrating the interaction between different product combinations.
	With multiple goods involved, the bundle-effects parameters directly are not readily interpretable.
	Therefore, the following subsection presents and discusses price elasticity, which determines the complementarity and substitutability among the goods.
	
	\begin{table}[hptb]
		\centering
		\caption{Estimates of Bundle-Effects Parameters}
		\resizebox{\textwidth}{!}{
			\footnotesize
\begin{tabular}{lcccc}
\hline
$j_1 =$ & 1 Sugary Soft Drinks & 1 Sugary Soft Drinks & 1 Sugary Soft Drinks & 1 Sugary Soft Drinks \\
$j_2 =$ & 2 Diet Soft Drinks & 3 Carbonated Water & 4 Milk Drinks & 5 Salty Snacks \bigstrut[b]\\
\hline
Constant & 0.472*** & -0.005 & 0.296** & 0.924*** \bigstrut[t]\\
      & (0.118) & (0.204) & (0.143) & (0.105) \\
Income & 0.045*** & 0.045** & 0.000 & -0.056*** \\
      & (0.011) & (0.020) & (0.014) & (0.010) \\
Child 0-5 & -0.032 & 0.175 & -0.018 & 0.002 \\
      & (0.063) & (0.116) & (0.054) & (0.050) \\
Child 6-11 & -0.050 & 0.008 & 0.061 & -0.043 \\
      & (0.040) & (0.080) & (0.041) & (0.034) \\
Child 12-17 & 0.015 & 0.018 & 0.004 & 0.040 \\
      & (0.035) & (0.065) & (0.036) & (0.029) \\
Family Size & 0.015 & -0.015 & -0.047*** & -0.010 \\
      & (0.013) & (0.023) & (0.013) & (0.010) \bigstrut[b]\\
\hline
$j_1 = $ & 2 Diet Soft Drinks & 2 Diet Soft Drinks & 2 Diet Soft Drinks &  \\
$j_2 = $ & 3 Milk Drinks & 4 Salty Snacks & 5 Carbonated Water &  \bigstrut[b]\\
\hline
Constant & 0.242 & 0.210* & 1.295*** &  \bigstrut[t]\\
      & (0.168) & (0.120) & (0.295) &  \\
Income & -0.011 & 0.005 & -0.076*** &  \\
      & (0.016) & (0.012) & (0.028) &  \\
Child 0-5 & -0.044 & -0.115** & -0.122 &  \\
      & (0.064) & (0.052) & (0.174) &  \\
Child 6-11 & -0.008 & -0.002 & 0.185** &  \\
      & (0.046) & (0.036) & (0.084) &  \\
Child 12-17 & -0.018 & -0.059* & -0.125 &  \\
      & (0.043) & (0.031) & (0.091) &  \\
Family Size & -0.002 & 0.002 & -0.035 &  \\
      & (0.016) & (0.010) & (0.022) &  \bigstrut[b]\\
\hline
$j_1 = $ & 3 Carbonated Water & 3 Carbonated Water &       & 4 Milk Drinks \\
$j_2 = $ & 4 Milk Drinks & 5 Salty Snacks &       & 5 Salty Snacks \bigstrut[b]\\
\hline
Constant & 0.217 & 0.079 &       & 0.303** \bigstrut[t]\\
      & (0.378) & (0.239) &       & (0.134) \\
Income & 0.003 & 0.008 &       & -0.009 \\
      & (0.034) & (0.023) &       & (0.013) \\
Child 0-5 & -0.177 & -0.296** &       & 0.003 \\
      & (0.125) & (0.127) &       & (0.054) \\
Child 6-11 & 0.200 & -0.097 &       & 0.019 \\
      & (0.118) & (0.085) &       & (0.037) \\
Child 12-17 & 0.098 & -0.096 &       & -0.022 \\
      & (0.101) & (0.064) &       & (0.034) \\
Family Size & -0.015 & 0.015 &       & -0.011 \\
      & (0.034) & (0.021) &       & (0.011) \bigstrut[b]\\
\hline
\multicolumn{5}{p{1.03\textwidth}}{Notes:
This table presents the posterior means and standard deviations (in parentheses) of the bundle-effects parameters from the main specification by the main model, the time-varying factor-augmented (TV-FA) bundle demand model with endogenous regressors.
Each column corresponds to a pair of goods, $j_1$ and $j_2$, illustrating the interaction between different product combinations. 
*** $p<0.01$, ** $p<0.05$, * $p<0.1$.
The significance levels are based on the posterior credibility intervals.
} \bigstrut[t]\\
\end{tabular}}
		\label{tab:soda_estimates_bundles}
	\end{table}

	\subsection{Estimates of Price Elasticities}
	
	\begin{table}[htb]
		\centering
		\caption{Estimates of Own- and Cross-Price Elasticities}
		\resizebox{\textwidth}{!}{
			\footnotesize
\begin{tabular}{llccccc}
\hline
 & \multicolumn{1}{r}{Good $k=$} & 1     & 2     & 3     & 4     & 5 \bigstrut[t]\\
Price $j=$ &       & Sugary Soft Drinks & Diet Soft Drinks & Carbonated Water & Milk Drinks & Salty Snacks \bigstrut[b]\\
\hline
1     & Sugary Soft Drinks & -0.668*** & -0.117*** & -0.054*** & -0.005 & -0.028*** \bigstrut[t]\\
      &       & (0.035) & (0.010) & (0.020) & (0.012) & (0.005) \\
2     & Diet Soft Drinks & -0.150*** & -0.770*** & -0.055*** & 0.002 & -0.028*** \\
      &       & (0.011) & (0.037) & (0.021) & (0.012) & (0.006) \\
3     & Carbonated Water & -0.006** & -0.005* & -0.500*** & 0.000 & -0.002 \\
      &       & (0.003) & (0.003) & (0.110) & (0.004) & (0.002) \\
4     & Milk Drinks & -0.003 & 0.003 & 0.001 & -1.145*** & -0.010** \\
      &       & (0.008) & (0.007) & (0.018) & (0.057) & (0.005) \\
5     & Salty Snacks & -0.160*** & -0.125*** & -0.086** & -0.066*** & -1.600*** \\
      &       & (0.018) & (0.017) & (0.042) & (0.024) & (0.034) \bigstrut[b]\\
\hline
\multicolumn{7}{p{1.2\textwidth}}{Notes:
This table presents the estimated price elasticities by the main model, the time-varying factor-augmented (TV-FA) bundle demand model with endogenous regressors (Endo).
Cell entries $(j,k)$, where $j$ indexes row and $k$ indexes column, give the posterior mean (and standard deviation in parentheses) of the percentage change in the probability of using $k$ substance in response to a percentage increase in the price of substance $j$.
*** $p<0.01$, ** $p<0.05$, * $p<0.1$.
The significance levels are based on the posterior credibility intervals.
} \bigstrut[t]\\
\end{tabular}}
		\label{tab:soda_elasticities_goods}
	\end{table}
	
	Table \ref{tab:soda_elasticities_goods} reports the estimated own- and cross-price elasticities of demand from the full model, specifically the time-varying factor-augmented (TV-FA) bundle demand model with endogenous regressors.
	Own-price elasticities, displayed along the diagonal, show the percentage change in the choice probability of good $k$ due to a one percent change in its price, holding other prices constant. 
	Aligned with expectation, all estimated own-price elasticities are negative.
	The statistically significant own-price elasticity of sugary soft drinks confirms that soda taxes decreases the intake of sugar from sugar-sweetened beverages, and consequently leads to the intended health benefits.
	
	The off-diagonal entries report cross-price elasticities, showing the percentage change in the choice probability of good $k$ (column) when the price of good $j$ (row) increases by one percent.
	Notably, the relationship between sugary soft drinks and milk drinks exhibits independence, as indicated by the insignificant cross-price elasticity $(j,k)=(1,4)$.
	This suggests that a soda tax does not drive consumers to milk drinks as an alternative source of sugar.
	Furthermore, sugary soft drinks and salty snacks are identified as weak complements, as shown by the significant but modest cross-price elasticity $(j,k)=(1,5)$.
	This suggests that implementing a soda tax could have ancillary health benefits by slightly reducing the consumption of salty snacks alongside sugary drinks, thereby contributing to a broader health benefit beyond the primary target of reducing sugar intake.

	\section{Conclusion}
	\label{sec:conclusion}
	
	This paper introduces a Bayesian factor-augmented bundle demand model for panel data.
	Its primary objective is to unravel the complexities of demand estimation, specifically focusing on the substitutability and complementarity among a variety of goods in the context of endogenous regressors.
	To tackle the challenges of taste correlation and endogeneity, the model employs a factor structure.
	Notably, this structure is enhanced with time-varying factor loadings, allowing it to adeptly capture the dynamic nature of unobserved tastes and utility-price confounders over time.
	
	The empirical results from these applications indicate significant price endogeneity and fluctuating unobserved tastes and utility-price confounders over time.
	An insight from this study is the tendency of beer products to demonstrate independent demand, as suggested by the statistically insignificant cross-price elasticities.
	This finding provides an understanding of consumer behavior in the beverage market. The research also highlights a consumer penchant for variety, evidenced by the escalating bundle effects correlating with family size.
	
	Extending upon Gentzkow’s model with complementarity, this research incorporates endogenous regressors into the framework.
	This is particularly pivotal given the often overlooked, yet crucial role of endogeneity in demand estimation.
	The factor approach introduced in this model is both flexible and parsimonious, effectively addressing the complexities brought about by endogeneity across both choices and time periods.
	
	The paper employs a multinomial probit (MNP) model with standard normal idiosyncratic errors to facilitate Bayesian posterior inference.
	The posterior inference is developed through a Markov-Chain Monte Carlo sampling scheme.
	The factor structure's comprehensive capture of error correlations streamlines the estimation process.
	Furthermore, vectorization of the model boosts the efficiency of sampling algorithms.
	
	Through extensive Monte Carlo simulation studies, the performance of the proposed model is scrutinized in comparison to benchmark models, including both time-invariant random-effects and factor-augmented models, with and without endogenous regressors.
	These studies confirm the model's ability to accurately recover true parameters and reliably estimate price elasticity.
	Importantly, they bring to light the crucial need to account for price endogeneity in complementarity estimation and the significance of incorporating time-varying unobserved tastes and utility-price confounders.
	
	The application to the sugary drink tax investigates concerns that such a tax might lead to an increased consumption of alternative sugary products and examines the potential health benefits from reducing the consumption of unhealthy snacks.
	The estimation of substitution patterns between sugary soft drinks, diet soft drinks, carbonated water, milk drinks, and salty snacks confirms that soda taxes decrease the intake of sugar from sugar-sweetened beverages, consequently leading to the intended health benefits.
	The results also suggest an independence between sugary soft drinks and milk drinks, implying that a soda tax does not cause consumers to substitute sugary soft drinks with milk drinks as an alternative source of sugar.
	Additionally, sugary soft drinks and salty snacks are identified as weak complements.
	This suggests that a soda tax could yield secondary health benefits by marginally decreasing the consumption of salty snacks alongside sugary drinks, thus extending the health benefits of the tax beyond merely reducing sugar intake alone.

	\bibliography{ref.bib}
	
	\newpage
	
	\appendix
	
	\section{Details on Model and Estimation}

	\subsection{Bayesian Probabilistic Model}
	\label{sec:appendix_model}
	
	This subsection presents the Bayesian probabilistic model introduced in Subsection \ref{sec:general_model}. To streamline notation, the symbol $p$ without subscript is used to denote discrete and continuous probabilistic functions, such as likelihood functions, prior and posterior densities. Meanwhile, $p_{ijt}$ with subscript refers specifically to the price of good $j$ at time $t$ faced by individual $i$.
	
	{\small
		\begin{align*}
			p(y_{it}=r^*|u_{it}) &=\mathbf{1}\left(u_{ir^*t}=\max_{r\in\mathcal{R}}\{u_{irt}\}\right), &&\text{for }t\in\mathcal{T}_i,i\in\mathcal{N}, \\
			u_{irt}|\theta,\gamma,\lambda,f_i,p_{it} &\sim\mathcal{N}\left(\mu_{irt}(\theta,\gamma)+\sum_{j\in r}\sum_{l=1}^L\lambda_{jlt}f_i,1\right), &&\text{for }r\in\mathcal{R}\backslash0,t\in\mathcal{T}_i,i\in\mathcal{N}, \\
			\mu_{irt}(\theta,\gamma) &=\sum_{j\in r}z_{ijt}'\theta_j+\sum_{j_1,j_2\in r}w_{i(j_1,j_2)t}'\gamma_{(j_1,j_2)}, &&\text{for }r\in\mathcal{R}\backslash0,t\in\mathcal{T}_i,i\in\mathcal{N}, \\
			u_{i0t} &\sim\mathcal{N}(0,1), &&\text{for }t\in\mathcal{T}_i,i\in\mathcal{N}, \\
			p_{ijt}|\theta^p_j,\lambda^p,f_i &\sim\mathcal{N}\left({z^p_{ijt}}'\theta^p_j+\sum_{l=1}^L\lambda^p_{jlt}f_i,1\right), &&\text{for }j\in\mathcal{J}_p,t\in\mathcal{T}_i,i\in\mathcal{N}, \\
			\Theta &\sim\mathcal{N}(m_\Theta,V_\Theta), &&\text{where }\Theta=\left(\theta',\gamma',{\theta^p}'\right)', \\
			\lambda_{jlt} &\sim\mathcal{N}(0,\sigma^2_\lambda), &&\text{for }j\in\mathcal{J},l=1,\ldots,L,t\in\mathcal{T}, \\
			\lambda^p_{jlt} &\sim\mathcal{N}(0,\sigma^2_\lambda), &&\text{for }j\in\mathcal{J}_p,l=1,\ldots,L,t\in\mathcal{T}, \\
			f_i &\sim\mathcal{N}(\mathbf{0},\mathbf{I}_L), &&\text{for }i\in\mathcal{N},
		\end{align*}
	}where $\mathbf{1}$ and $\mathbf{I}$ denotes the indicator function and an identity matrix with dimension specified in subscript respectively.
	Data contain observed bundle choice $\{\{y_{it}\}_{t\in\mathcal{T}_i}\}_{i\in\mathcal{N}}$, endogenous regressors $\{\{\{p_{ijt}\}_{j\in\mathcal{J}_p}\}_{t\in\mathcal{T}_i}\}_{i\in\mathcal{N}}$, utility exogenous and endogenous regressors $\{\{\{z_{ijt}\}_{j\in\mathcal{J}}\}_{t\in\mathcal{T}_i}\}_{i\in\mathcal{N}}$, bundle-effects regressors $\{\{\{w_{i(j_1,j_2)t}\}_{j_1,j_2\in\mathcal{J}}\}_{t\in\mathcal{T}_i}\}_{i\in\mathcal{N}}$, instruments $\{\{\{z^p_{ijt}\}_{j\in\mathcal{J}_p}\}_{t\in\mathcal{T}_i}\}_{i\in\mathcal{N}}$. Latent utilities $\{\{\{u_{irt}\}_{r\in\mathcal{R}}\}_{t\in\mathcal{T}_i}\}_{i\in\mathcal{N}}$ are augmented data and in parameter space.
	Other model parameters are parameter vector $\Theta=\left(\theta',\gamma',{\theta^p}'\right)'$, preference parameters $\theta=\left(\{\theta_j\}_{j\in\mathcal{J}}\right)'$, bundle-effects parameters $\gamma=\left(\{\gamma_{(j_1,j_2)}\}_{j_1,j_2\in\mathcal{J}}\right)'$, first-stage parameters $\theta^p=\left(\{\theta^p_j\}_{j\in\mathcal{J}_p}\right)'$, non-zero loadings $\lambda=\left(\{\{\{\lambda_{jlt}\}_{j\in\mathcal{J}},\{\lambda^p_{jlt}\}_{j\in\mathcal{J}_p}\}_{l=1}^L\}_{t\in\mathcal{T}}\right)'$, and factors $f=\left(\{f_i\}_{i\in\mathcal{N}}\right)'$.
	Hyperparameters are $m_\Theta$, $V_\Theta$, $\sigma^2_\lambda$.
	
	To allow sparsity implied by prior knowledge or economic theory, the structure of factor loading matrix $\{\Lambda_t\}_{t\in\mathcal{T}}$ is governed by an indicator matrix $\delta$.
	$\lambda_{jlt}=0$ (or $\lambda^p_{jlt}=0$) if $\delta_{jlt}=0$, and $\lambda_{jlt}\sim\mathcal{N}(0,\sigma^2_\lambda)$ if $\delta_{jlt}=1$.

	\subsection{Model Vectorization}
	\label{sec:appendix_vectorization}
	
	This subsection provides details on model vectorization with an example.
	Consider there are $J=3$ goods, and the $J_p=3$ good prices $p_{ijt}$ are endogenous.
	Specify the good utility and first-stage equations \eqref{eq:general_utility_j} and \eqref{eq:general_price} as:
	\begin{align*}
		\bar{u}_{ijt}&=\alpha p_{ijt}+\beta_{0j}+\beta_{1j}x_{ijt}+\nu_{ijt}, && \text{for } j\in\mathcal{J},t\in\mathcal{T}_i,i\in\mathcal{N}, \\
		p_{ijt}&={z_{ijt}^p}'\theta^p_j+\nu^p_{ijt}+\epsilon_{ijt}^p, && \text{for } j\in\mathcal{J}_p,t\in\mathcal{T}_i,i\in\mathcal{N}.
	\end{align*}
	To be consistent, define the covariate and parameters vectors as $z_{ijt}=(p_{ijt},1,x_{ijt})'$, $\theta_j=(\alpha,\beta_{0j},\beta_{1j})'$. Note that the price coefficient $\alpha$ is a common parameter for all prices $p_{ijt}$. Similarly, specify the bundle effects in Equation \eqref{eq:general_utility_r_2} as
	\begin{equation*}
		\Gamma_{i(j_1,j_2)t} =\tilde{\gamma}\tilde{w}_{i(j_1,j_2)t}+\tilde{\gamma}_{(j_1,j_2)},
	\end{equation*}
	with $w_{i(j_1,j_2)t}=(\tilde{w}_{i(j_1,j_2)t},1)'$ and $\gamma_{(j_1,j_2)}=(\tilde{\gamma},\tilde{\gamma}_{(j_1,j_2)})'$. Note that $\tilde{\gamma}$ is a common parameter to all bundle effects.
	
	Specify the order of the choice set (bundles)
	\begin{align*}
		\mathcal{R}=\Big\{&0 \text{ (or $\emptyset$)}, \nonumber\\
		&1,2,\ldots,J, \nonumber\\
		&(1,2),\ldots,(1,J),(2,3),\ldots,(2,J),\ldots,(J-1,J), \nonumber\\
		&(1,2,3),\ldots,(1,2,J),(1,3,4),\ldots,(1,3,J),\ldots,(J-2,J-1,J), \nonumber\\
		&\ldots, \nonumber\\
		&(1,2,\ldots,J)\Big\},
	\end{align*}
	according to the ordered product set $\mathcal{J}=\{1,2,\ldots,J\}$, where $J=|\mathcal{J}|$ is the number of product.
	Define the number of inside options to be $R=|\mathcal{R}|-1=2^J-1$, and express the full choice set as an ordered set, $\mathcal{R}=\{0,1,2,\ldots,R\}$. In this example with $J=3$ goods, there are $R=7$ inside options:
	\begin{equation*}
		\mathcal{R}=\big\{0,1,2,3,(1,2),(1,3),(2,3),(1,2,3)\big\}.
	\end{equation*}
	
	Stack bundle utilities \eqref{eq:general_utility_r_2} over bundles $r\in\mathcal{R}\backslash0$ according to the specified order and obtain Equation \eqref{eq:utility_it}:
	\begin{equation*}
		u_{it}=z_{it}\theta+w_{it}\gamma+I_\nu^u\nu_{it}^u+\epsilon_{it}^u, \qquad\text{for }t\in\mathcal{T}_i,i\in\mathcal{N},
	\end{equation*}
	where
	\begin{align*}
		u_{it}&=(u_{i1t},u_{i2t},u_{i3t},u_{i(1,2)t},u_{i(1,3)t},u_{i(2,3)t},u_{i(1,2,3)t})', \\
		\theta&=(\theta_1',\theta_2',\theta_3')'=(\alpha,\beta_{01},\beta_{11},\beta_{02},\beta_{12},\beta_{03},\beta_{13})', \\
		\gamma&=(\gamma_{(1,2)}',\gamma_{(1,3)}',\gamma_{(2,3)}')'=(\tilde{\gamma},\tilde{\gamma}_{(1,2)},\tilde{\gamma}_{(1,3)},\tilde{\gamma}_{(2,3)})', \\
		\nu_{it}^u&=(\nu_{i1t},\nu_{i2t},\nu_{i3t})', \\
		\epsilon^u_{it}&=(\epsilon_{i1t},\epsilon_{i2t},\epsilon_{i3t},\epsilon_{i(1,2)t},\epsilon_{i(1,3)t},\epsilon_{i(2,3)t},\epsilon_{i(1,2,3)t})',
	\end{align*}
	and vectorized covariate and mapping matrices are
	\begin{equation*}
		z_{it}=\begin{pmatrix}
			p_{i1t}&1&x_{i1t}&0&0&0&0 \\
			p_{i2t}&0&0&1&x_{i2t}&0&0 \\
			p_{i3t}&0&0&0&0&1&x_{i3t} \\
			p_{i1t}+p_{i2t}&1&x_{i1t}&1&x_{i2t}&0&0 \\
			p_{i1t}+p_{i2t}&1&x_{i1t}&0&0&1&x_{i3t} \\
			p_{i1t}+p_{i2t}&0&0&1&x_{i2t}&1&x_{i3t} \\
			p_{i1t}+p_{i2t}+p_{i3t}&1&x_{i1t}&1&x_{i2t}&1&x_{i3t}
		\end{pmatrix},
	\end{equation*}
	\begin{equation*}
		w_{it}=\begin{pmatrix}
			0&0&0&0 \\
			0&0&0&0 \\
			0&0&0&0 \\
			\tilde{w}_{i(1,2)t}&1&0&0 \\
			\tilde{w}_{i(1,3)t}&0&1&0 \\
			\tilde{w}_{i(2,3)t}&1&0&1 \\
			\tilde{w}_{i(1,2)t}+\tilde{w}_{i(1,3)t}+\tilde{w}_{i(2,3)t}&1&1&1
		\end{pmatrix}, \qquad
		I_\nu^u=\begin{pmatrix}
			1&0&0 \\
			0&1&0 \\
			0&0&1 \\
			1&1&0 \\
			1&0&1 \\
			0&1&1 \\
			1&1&1
		\end{pmatrix}.
	\end{equation*}
	
	Similarly, stack first-stage equations \eqref{eq:general_price} over $j\in\mathcal{J}_p=(1,2,3)$ and obtain Equation \eqref{eq:price_it}:
	\begin{equation*}
		p_{it}=z_{it}^p\theta^p+I_\nu^p\nu_{it}^p+\epsilon_{it}^p, \qquad\text{for }t\in\mathcal{T}_i,i\in\mathcal{N},
	\end{equation*}
	where
	\begin{align*}
		p_{it}&=(p_{i1t},p_{i2t},p_{i3t})', \\
		\theta^p&=({\theta^p_1}',{\theta^p_2}',{\theta^p_3}')', \\
		\nu^p_{it}&=(\nu^p_{i1t},\nu^p_{i2t},\nu^p_{i3t})', \\
		\epsilon_{it}^p&=(\epsilon_{i1t}^p,\epsilon_{i2t}^p,\epsilon_{i3t}^p)',
	\end{align*}
	and vectorized mapping matrix $I_\nu^p=\mathbf{I}_3$ and covariate matrix
	\begin{equation*}
		z_{it}^p=\begin{pmatrix}
			z_{i1t}'&\mathbf{0}&\mathbf{0} \\
			\mathbf{0}&z_{i2t}'&\mathbf{0} \\
			\mathbf{0}&\mathbf{0}&z_{i3t}'
		\end{pmatrix}.
	\end{equation*}
	
	Stack the latent utilities and endogenous regressors $y_{it}^*=(u_{it}',p_{it}')'$ and obtain Equation \eqref{eq:y_star_RE}:
	\begin{equation*}
		y_{it}^*=h_{it}\Theta+I_\nu\nu_{it}+\epsilon_{it}, \qquad\text{ for }t\in\mathcal{T}_i,i\in\mathcal{N},
	\end{equation*}
	where
	\begin{align*}
		\Theta&=(\theta',\gamma',{\theta^p}')', \\
		\nu_{it}&=({\nu_{it}^u}',{\nu_{it}^p}')', \\
		\epsilon_{it}&=({\epsilon_{it}^u}',{\epsilon_{it}^p}')',
	\end{align*}
	and vectorized covariate and mapping matrices
	\begin{equation*}
		h_{it}=\begin{pmatrix}
			z_{it}&w_{it}&\mathbf{0} \\
			\mathbf{0}&\mathbf{0}&z_{it}^p
		\end{pmatrix}, \qquad
		I_\nu=\begin{pmatrix}
			I_\nu^u&\mathbf{0} \\
			\mathbf{0}&I_\nu^p
		\end{pmatrix}.
	\end{equation*}
	
	Augment the structural error vector with the factor structure $\nu_{it}=({\nu_{it}^u}',{\nu_{it}^p}')'=\Lambda_t f_i$ and obtain Equation \eqref{eq:y_star_FA}:
	\begin{equation*}
		y_{it}^*=h_{it}\Theta+I_\nu\Lambda_t f_i+\epsilon_{it}, \qquad\text{ for }t\in\mathcal{T}_i,i\in\mathcal{N}, 
	\end{equation*}
	where $\Lambda_t$ represents the $(J+J_p)$-by-$L$ factor loading matrix for $t\in\mathcal{T}$ and $f_i$ the $L$-by-$1$ latent factors for $i\in\mathcal{N}$.
	
	Stack over $t\in\mathcal{T}_i$ for each individual $i$ and obtain Equation \eqref{eq:y_star_i}: 
	\begin{equation*}
		y_i^*=h_i\Theta+I_{\nu,i}^T\Lambda f_i+\epsilon_i, \qquad\text{ for }i\in\mathcal{N},
	\end{equation*}
	where $h_i=(\{h_{it}'\}_{t\in\mathcal{T}_i})'$.
	$I_{\nu,i}^T$ is block matrix with $T_i$-by-$T$ blocks.
	For each block $t_1\in\mathcal{T}_i$ and $t_2\in\mathcal{T}$, $I_{\nu,i}^T(t_1,t_2)=I_\nu$ if $t_1=t_2$, $I_{\nu,i}^T(t_1,t_2)=\mathbf{0}$ otherwise.
	If the data is a balanced panel, i.e. $\mathcal{T}_i=\mathcal{T}$, then
	\begin{equation*}
		I_\nu^T=\begin{pmatrix}
			I_\nu&\mathbf{0}&\cdots&\mathbf{0} \\
			\mathbf{0}&I_\nu&\cdots&\mathbf{0} \\
			\vdots&\vdots&\ddots&\vdots \\
			\mathbf{0}&\mathbf{0}&\cdots&I_\nu
		\end{pmatrix}
	\end{equation*}
	The collected factor loading matrix $\Lambda=(\Lambda_1',\ldots,\Lambda_T)'$ is $(J+J_p)T$-by-$L$.
	In the case of time-invariant factor loading, the collected factor loading matrix $\Lambda$ has a dimension of $\Lambda$ $(J+J_p)$-by-$L$, and $I_{\nu,i}^T=(I_\nu',\ldots,I_\nu')'$ stacking $T_i$ $I_\nu$'s vertically. 
	
	Lastly, stack over $i\in\mathcal{N}$ and obtain Equations \eqref{eq:y_star_1} and \eqref{eq:y_star_2}:
	\begin{align*}
		y^*&=H_{zw}\Theta+H_\Lambda f+\epsilon, \\
		&=H_{zw}\Theta+H_f\lambda+\epsilon,
	\end{align*}
	where
	\begin{align*}
		y^*&=({y_1^*}',\ldots,{y_N^*}')', \\
		H_{zw}&=(h_1',\ldots,h_N')', \\
		f&=(f_1',\ldots,f_N')', \\
		\lambda&=\left(\{\{\{\lambda_{jlt}\}_{j\in\mathcal{J}|\delta_{jlt}=1},\{\lambda^p_{jlt}\}_{j\in\mathcal{J}_p|\delta_{jlt}=1}\}_{l=1}^L\}_{t\in\mathcal{T}}\right)', \\
		\epsilon&=(\epsilon_1',\ldots,\epsilon_N')',
	\end{align*} 
	\begin{equation*}
		H_\Lambda=\begin{pmatrix}
			I_{\nu,1}^T\Lambda&\mathbf{0}&\cdots&\mathbf{0} \\
			\mathbf{0}&I_{\nu,2}^T\Lambda&\cdots&\mathbf{0} \\
			\vdots&\vdots&\ddots&\vdots \\
			\mathbf{0}&\mathbf{0}&\cdots&I_{\nu,N}^T\Lambda
		\end{pmatrix},
	\end{equation*}
	and $H_f$ is $N$-by-$L$ block matrix. For each block $i=1,\ldots,N$ and $l=1,\ldots,L$, $H_f(i,l)=f_{il}I_{\nu,i}^{T,l}$.
	Columns of $I_{\nu,i}^{T}$ correspond to
	\begin{equation*}
		(j,t) = (1,1), \ldots, (J,1), (J+1,1), \ldots, (J+J_p,1), (1,2), \ldots,(J+J_p,T),
	\end{equation*}
	and for each $l=1,\ldots,L$, $I_{\nu,i}^{T,l}$ includes the columns of $I_{\nu,i}^{T}(j,t)$ that $\delta_{jlt}=1$.

	\subsection{Joint Error Vector and Its Covariance}
	\label{sec:appendix_joint_error}
	
	The process of model vectorization defines joint error vectors at different levels of aggregation. In Equation \eqref{eq:y_star_FA},
	\begin{equation*}
		\varepsilon_{it}=I_\nu\Lambda_t f_i+\epsilon_{it}\sim\mathcal{N}(\mathbf{0},\Omega_{t}), \qquad\text{for }t\in\mathcal{T}_i, i\in\mathcal{N},
	\end{equation*}
	with covariance decomposition ($J+J_p$-by-$J+J_p$)
	\begin{equation*}
		\Omega_{t}=\mathrm{cov}(\varepsilon_{it})=I_\nu\Lambda_t\Lambda_t'I_\nu'+\mathbf{I},
	\end{equation*}
	The covariance parameters in $\Omega_{t}$ capture error correlations between latent utilities (taste correlations) and between utility and endogenous regressors (endogeneity) within each time period $t\in\mathcal{T}$.
	
	In equation \eqref{eq:y_star_i},
	\begin{equation*}
		\varepsilon_i=I_{\nu,i}^T\Lambda f_i+\epsilon_i\sim\mathcal{N}(\mathbf{0},\Omega_i), \qquad\text{for }i\in\mathcal{N},
	\end{equation*}
	with covariance decomposition ($(J+J_p)T_i$-by-$(J+J_p)T_i$)
	\begin{equation*}
		\Omega_i=\mathrm{cov}(\varepsilon_i)=I_{\nu,i}^T\Lambda\Lambda'{I_{\nu,i}^T}'+\mathbf{I},
	\end{equation*}
	which further captures intertemporal error correlations.
	
	Lastly, in equation \eqref{eq:y_star_1} and \eqref{eq:y_star_2},
	\begin{equation*}
		\varepsilon=H_\Lambda f+\epsilon=H_f\lambda+\epsilon\sim\mathcal{N}(\mathbf{0},\Omega),
	\end{equation*}
	where
	\begin{equation*}
		\Omega=\begin{pmatrix}
			\Omega_1&\mathbf{0}&\cdots&\mathbf{0} \\
			\mathbf{0}&\Omega_2&\cdots&\mathbf{0} \\
			\vdots&\vdots&\ddots&\vdots \\
			\mathbf{0}&\mathbf{0}&\cdots&\Omega_N
		\end{pmatrix}.
	\end{equation*}

	\subsection{MCMC Sampling Scheme}
	\label{sec:appendix_mcmc}
	
	Rewrite the Bayesian probabilistic model presented in Subsection \ref{sec:appendix_model} in vectorized form.
	\begin{align*}
		p(y_{it}=r^*|u_{it}) &=\mathbf{1}\left(u_{ir^*t}=\max_{r\in\mathcal{R}}\{u_{irt}\}\right), \qquad\text{for }t\in\mathcal{T}_i,i\in\mathcal{N}, \\
		y^*|\Theta,\lambda,f&\sim\mathcal{N}(H_{zw}\Theta+H_\Lambda f,\mathbf{I}), \\
		\Theta&\sim\mathcal{N}(m_\Theta,V_\Theta), \\
		f&\sim\mathcal{N}(\mathbf{0},\mathbf{I}_{N\times L}), \\
		\lambda&\sim\mathcal{N}(\mathbf{0},\sigma^2_\lambda\mathbf{I}_{\sum\mathbf{1}(\delta_{jlt}=1)}).
	\end{align*}
	Note that $y^*$ has three parameterization that facilitate the sampling of $\Theta$, $\lambda$ and $f$:
	\begin{align*}
		y^*|\Theta,\lambda,f&\sim\mathcal{N}(H_{zw}\Theta+H_f\lambda,\mathbf{I}), \\
		y^*|\Theta,\lambda,f&\sim\mathcal{N}(H_{zw}\Theta+H_\Lambda f,\mathbf{I}), \\
		y^*|\Theta,\lambda&\sim\mathcal{N}(H_{zw}\Theta,\Omega).
	\end{align*}
	I acknowledge that the joint densities of latent utilities $u_{irt}$ and endogenous regressors $p_{ijt}$ are not proper likelihood functions, as the endogenous regressors $p_{ijt}$ present on both sides, in $y^*$ and $H_{zw}$. Here, consider the densities as functions of model parameters $\Theta$, $\Lambda$ and $f$. These densities in joint form are convenient when deriving (full conditional) posterior densities.
	
	I employ the following Markov-Chain Monte Carlo sampling scheme to generate draws from the joint posterior distribution of the model parameters.
	\begin{itemize}
		\item[1.] For $i\in\mathcal{N}$, $t\in\mathcal{T}_i$ and $r\in\mathcal{R}$, sample the latent utility from a truncated normal distribution
		\begin{equation*}
			u_{irt}|u_{i,-r,t},\Theta,\lambda,f_i\sim\begin{cases}
				\mathcal{N}(\mu(u_{irt}),1)\mathbf{1}(u_{irt}>\max(u_{i,-r,t})),\text{ if }y_{it}=r, \\
				\mathcal{N}(\mu(u_{irt}),1)\mathbf{1}(u_{irt}<\max(u_{i,-r,t})),\text{ if }y_{it}\neq r,
			\end{cases}
		\end{equation*}
		where $u_{i,-r,t}$ denotes the latent utilities in the choice set $\mathcal{R}$ other than $r$, and
		\begin{equation*}
			\mu(u_{irt})=\begin{cases}\sum_{j\in r}z_{ijt}'\theta_j+\sum_{j_1,j_2\in r}w_{i(j_1,j_2)t}'\gamma_{(j_1,j_2)}+\sum_{j\in r}\sum_{l=1}^L\lambda_{jlt}f_i, & \text{for }r\in\mathcal{R}\setminus0, \\
				0, & \text{for }r=0.
			\end{cases}
		\end{equation*}
		
		\item[2.] Sample the equation parameters $\Theta=(\theta',\gamma',{\theta^p}')'$ from
		\begin{equation*}
			\Theta|u,\lambda,f\sim\mathcal{N}(\bar{m}_\Theta,\bar{V}_\Theta),
		\end{equation*}
		where $\bar{V}_\Theta=(H_{zw}'H_{zw}+V_\Theta^{-1})^{-1}$, $\bar{m}_\Theta=\bar{V}_\Theta(H_{zw}'\tilde{y}^*+V_\Theta^{-1}m_\Theta)$ and $\tilde{y}^*=y^*-H_\Lambda f$. \\
		
		An alternative sampler involves partially marginalizing over the factor $f$, and works with $y^*|\Theta,\lambda\sim\mathcal{N}(H_{zw}\Theta,\Omega)$ in the spirit of \cite{wagner2023factor}.
		Then, the equation parameter vector $\Theta=(\theta',\gamma',{\theta^p}')'$ is sampled from conditional posterior
		\begin{equation*}
			\Theta|u,\lambda\sim\mathcal{N}(\bar{m}_\Theta,\bar{V}_\Theta),
		\end{equation*}
		where $\bar{V}_\Theta=(H_{zw}'\Omega^{-1}H_{zw}+V_\Theta^{-1})^{-1}$, $\bar{m}_\Theta=\bar{V}_\Theta(H_{zw}'\Omega^{-1}y^*+V_\Theta^{-1}m_\Theta)$.
		This sampler is generally more efficient.
		However, it is only applicable when $\Omega$ has low dimensions and relatively sparse.
		$\Omega_i$ for $i\in\mathcal{N}$ are the non-zero main diagonal block matrices of $\Omega$, and has dimensions of $(R+J_p)T_i$-by-$(R+J_p)T_i$.
		As $N$, $J$, $R=2^J-1$, $J_p$ and $T$ increase, $\Omega$ increases in its size and is less sparse.
		Consequently, the alternative sampler with algorithm efficiency requires longer computational time.
		
		\item[3.] Sample latent factors $f=(f_1',\ldots,f_N')'$ from
		\begin{equation*}
			f|u,\Theta,\lambda\sim\mathcal{N}(\bar{m}_f,\bar{V}_f),
		\end{equation*}
		where $\bar{V}_f=(H_\Lambda'H_\Lambda+\mathbf{I})^{-1}$, $\bar{m}_f=\bar{V}_f H_\Lambda'\tilde{\tilde{y}}^*$, and $\tilde{\tilde{y}}^*=y^*-H_{zw}\Theta$.
		
		\item[4.] Random sign switch of $f_l$ and $\lambda_l$ for $l=1,\ldots,L$ \citep{jacobi2016bayesian}.
		
		\item[5.] Boosting MCMC with marginal data augmentation \citep{van2001art}: for $l=1,\ldots,L$, 
		\begin{itemize}
			\item[(a)] sample working parameter $\Psi_l\sim\mathcal{GIG}(p_\Psi,a_\Psi,b_\Psi)$ from generalized inverse Gaussian distribution,
			\item[(b)] update $\Psi_l^{\mathrm{new}}\sim\mathcal{GIG}(\bar{p}_{\Psi l},\bar{a}_{\Psi l},\bar{b}_{\Psi l})$, where $\bar{p}_{\Psi l}=p_\Psi+(\sum_{t\in\mathcal{T}}\sum_{j\in\mathcal{J}\cup\mathcal{J}_p}\delta_{jlt})/2-N/2$, $\bar{a}_{\Psi l}=a_\Psi+(\lambda_l'\lambda_l)/(\Psi_l\sigma^2_\lambda)$, $\bar{b}_{\Psi l}=b_\Psi+\Psi_l f_l' f_l$,
			\item[(c)] update $\lambda_l^{\mathrm{new}}=\lambda_l\sqrt{\Psi_l^{\mathrm{new}}/\Psi_l}$ and $f_l^{\mathrm{new}}=f_l\sqrt{\Psi_l/\Psi_l^{\mathrm{new}}}$.
		\end{itemize}
		\item[6.] Sample non-zero factor loadings $\lambda=\left(\{\{\{\lambda_{jlt}\}_{j\in\mathcal{J}|\delta_{jlt}=1},\{\lambda^p_{jlt}\}_{j\in\mathcal{J}_p|\delta_{jlt}=1}\}_{l=1}^L\}_{t\in\mathcal{T}}\right)'$ in $\Lambda$ from
		\begin{equation*}
			\lambda|u,\Theta,f\sim\mathcal{N}(\bar{m}_\lambda,\bar{V}_\lambda),
		\end{equation*}
		where $\bar{V}_\lambda=(H_f'H_f+\sigma^{-2}_\lambda\mathbf{I})^{-1}$, $\bar{m}_\lambda=\bar{V}_\lambda H_f'\tilde{\tilde{y}}^*$, and $\tilde{\tilde{y}}^*=y^*-H_{zw}\Theta$.
	\end{itemize}

	
	\section{Supplementary Results}
	
	\subsection{Estimates of the First-Stage Equations for the Main Model}
	\label{sec:appendix_sup_first_stage}
	
	\begin{table}[!h]
		\centering
		\caption{Estimates of the First-Stage Equations for the Main Model}
		\resizebox{\textwidth}{!}{
			\footnotesize
			\begin{tabular}{llcccccc}
\hline
      & \multicolumn{1}{r}{$j = $} & 1     & 2     & 3     & 4     & 5     & 6 \bigstrut[t]\\
      &       & AB InBev & Boston Beer & Heineken & MillerCoors & Other Vendors & Malt Beverage \bigstrut[b]\\
\hline
\multicolumn{8}{l}{\boldmath{}\textbf{First-Stage Equation $p_{ijt}$}\unboldmath{}} \bigstrut[t]\\
      & Constant & -3.009 & 4.597*** & -3.666 & -0.761 & -28.797*** & 4.024 \\
      &       & (2.652) & (1.546) & (2.601) & (5.138) & (3.399) & (2.644) \\
      & No. of Shopping Trips & 0.000 & 0.000 & 0.000 & 0.000 & 0.000 & 0.000 \\
      &       & (0.001) & (0.001) & (0.001) & (0.001) & (0.001) & (0.001) \\
      & Household Income & 0.000 & 0.000 & 0.000 & 0.000 & 0.000 & 0.000 \\
      &       & (0.009) & (0.009) & (0.009) & (0.009) & (0.009) & (0.009) \\
      & Cabled TV & 0.000 & 0.000 & 0.000 & 0.000 & 0.000 & 0.000 \\
      &       & (0.020) & (0.019) & (0.019) & (0.020) & (0.019) & (0.019) \\
      & Pittsfield & 0.052 & -4.706* & 5.941 & 12.248 & 22.082*** & -1.450 \\
      &       & (4.564) & (2.625) & (4.463) & (8.674) & (5.818) & (4.518) \\
      & Family Size (Demeaned) & 0.000 & 0.000 & 0.000 & 0.000 & 0.000 & 0.000 \\
      &       & (0.006) & (0.006) & (0.006) & (0.006) & (0.006) & (0.006) \\
      & Price Instrument City 1 & 0.233 & -0.439*** & 0.060 & -0.210 & 3.922*** & -0.518*** \\
      &       & (0.481) & (0.166) & (0.143) & (0.317) & (0.609) & (0.161) \\
      & Price Instrument City 2 & -0.111 & -0.254*** & -0.393*** & 0.073 & 2.459*** & -0.421** \\
      &       & (0.345) & (0.099) & (0.083) & (0.134) & (0.458) & (0.164) \\
      & Price Instrument City 3 & 0.081 & 0.427*** & 1.281*** & 0.362 & -4.008*** & 0.104 \\
      &       & (0.377) & (0.169) & (0.176) & (0.381) & (0.670) & (0.240) \\
      & Price Instrument City 4 & 0.084 & 0.086 & 0.220 & -0.085 & 2.685*** & 0.237 \\
      &       & (0.354) & (0.085) & (0.179) & (0.167) & (0.342) & (0.349) \\
      & Price Instrument City 5 & 0.786 & -0.467 & 0.977*** & 0.104 & -1.365*** & 0.924*** \\
      &       & (0.703) & (0.452) & (0.188) & (0.588) & (0.500) & (0.178) \\
      & Price Instrument City 6 & 0.001 & -0.410*** & -0.790*** & -0.085 & -0.273*** & 0.284* \\
      &       & (0.355) & (0.153) & (0.171) & (0.222) & (0.085) & (0.166) \\
      & Price Instrument City 7 & 0.487 & 0.597 & -0.354 & 0.536 & 2.812*** & 0.659*** \\
      &       & (0.464) & (0.396) & (0.255) & (0.372) & (0.479) & (0.243) \\
      & Price Instrument City 8 & 0.245 & 0.095* & 0.434 & 0.357 & -0.936*** & -1.126*** \\
      &       & (0.462) & (0.049) & (0.340) & (0.345) & (0.133) & (0.254) \\
      & Price Instrument City 9 & -0.100 & 0.701*** & -0.036 & 0.039 & -0.081 & 0.084 \\
      &       & (0.116) & (0.165) & (0.100) & (0.258) & (0.097) & (0.124) \\
      & Price Instrument City 1 $\times$ Pittsfield & 1.367* & 3.071*** & -0.063 & 0.356 & -3.980*** & 0.973*** \\
      &       & (0.819) & (0.287) & (0.246) & (0.541) & (1.042) & (0.272) \\
      & Price Instrument City 2 $\times$ Pittsfield & -0.716 & 1.330*** & 0.587*** & -0.699*** & -1.314* & 0.037 \\
      &       & (0.592) & (0.170) & (0.143) & (0.230) & (0.780) & (0.279) \\
      & Price Instrument City 3 $\times$ Pittsfield & 0.765 & -2.461*** & -1.365*** & -0.925 & 2.264** & 0.297 \\
      &       & (0.639) & (0.290) & (0.299) & (0.647) & (1.143) & (0.413) \\
      & Price Instrument City 4 $\times$ Pittsfield & 1.075* & -0.960*** & -0.461 & -0.598** & -1.251** & -0.351 \\
      &       & (0.605) & (0.143) & (0.309) & (0.284) & (0.584) & (0.599) \\
      & Price Instrument City 5 $\times$ Pittsfield & -1.352 & 4.588*** & -0.783** & -1.670* & 0.279 & -1.333*** \\
      &       & (1.200) & (0.771) & (0.320) & (0.998) & (0.846) & (0.302) \\
      & Price Instrument City 6 $\times$ Pittsfield & -0.690 & 1.922*** & 0.779*** & 0.602 & 0.627*** & 0.232 \\
      &       & (0.603) & (0.259) & (0.292) & (0.380) & (0.146) & (0.283) \\
      & Price Instrument City 7 $\times$ Pittsfield & 1.151 & -3.626*** & 0.266 & 0.196 & -0.326 & -0.423 \\
      &       & (0.791) & (0.677) & (0.438) & (0.640) & (0.813) & (0.415) \\
      & Price Instrument City 8 $\times$ Pittsfield & -1.510* & -0.427*** & -0.613 & -1.246** & 0.597*** & 1.317*** \\
      &       & (0.785) & (0.083) & (0.583) & (0.589) & (0.227) & (0.435) \\
      & Price Instrument City 9 $\times$ Pittsfield & -0.244 & -2.735*** & 0.869*** & 1.226*** & 0.163 & -0.307 \\
      &       & (0.195) & (0.280) & (0.170) & (0.444) & (0.166) & (0.213) \bigstrut[b]\\
\hline
\multicolumn{8}{p{1.3\textwidth}}{Notes:
This table presents the posterior means and standard deviations (in parentheses) of the first-stage parameters from the main specification, introduced in Subsection \ref{sec:data_and_specification}, and by the main model, the time-varying factor-augmented (TV-FA) bundle demand model with endogenous regressors.
*** $p<0.01$, ** $p<0.05$, * $p<0.1$. The significance levels are based on the posterior credibility intervals.
The six columns correspond to the six goods.
} \bigstrut[t]\\
\end{tabular}
}
		\label{tab:first_stage}
	\end{table}
	
	\newpage
	\subsection{Estimates for the Benchmark Models for Comparison}
	\label{sec:appendix_sup_estimates}
	
	\begin{table}[!h]
		\centering
		\caption{Estimates for the Exogenous RE Bundle Demand Model}
		\resizebox{\textwidth}{!}{
			\footnotesize
			\begin{tabular}{llcccccc}
\hline
      & \multicolumn{1}{r}{$j = $} & 1     & 2     & 3     & 4     & 5     & 6 \bigstrut[t]\\
      &       & AB InBev & Boston Beer & Heineken & MillerCoors & Other Vendors & Malt Beverage \bigstrut[b]\\
\hline
\multicolumn{8}{l}{\boldmath{}\textbf{Good Utility $\bar{u}_{ijt}$}\unboldmath{}} \bigstrut[t]\\
      & Price & \multicolumn{6}{c}{-0.246***} \\
      &       & \multicolumn{6}{c}{(0.076)} \\
      & Constant & -2.797*** & -5.302*** & -6.394*** & -1.829*** & -2.452*** & -1.962*** \\
      &       & (0.677) & (0.921) & (1.205) & (0.669) & (0.642) & (0.716) \\
      & No. of Shopping Trips & 0.046*** & 0.034*** & 0.025*** & 0.053*** & 0.055*** & 0.019*** \\
      &       & (0.005) & (0.007) & (0.009) & (0.005) & (0.004) & (0.006) \\
      & Household Income & 0.076 & 0.220*** & 0.283** & 0.026 & 0.147*** & 0.032 \\
      &       & (0.055) & (0.061) & (0.108) & (0.054) & (0.046) & (0.045) \\
      & Cabled TV & -0.119 & 0.238 & 0.413** & 0.256** & 0.139 & 0.126 \\
      &       & (0.125) & (0.183) & (0.175) & (0.100) & (0.095) & (0.102) \\
      & Pittsfield & 0.535*** & 1.185*** & 0.721*** & -0.820*** & -0.364*** & 0.404*** \\
      &       & (0.113) & (0.118) & (0.162) & (0.124) & (0.126) & (0.102) \bigstrut[b]\\
\hline
\multicolumn{8}{l}{\boldmath{}\textbf{Bundle-Effects $\Gamma_{i(j_1,j_2)t}$}\unboldmath{}} \bigstrut[t]\\
      & Family Size (Demeaned) & \multicolumn{6}{c}{0.020*} \\
      &       & \multicolumn{6}{c}{(0.011)} \\
      & Boston Beer & 0.564*** &       &       &       &       &  \\
      &       & (0.114) &       &       &       &       &  \\
      & Heineken USA & 0.285** & 0.450*** &       &       &       &  \\
      &       & (0.119) & (0.142) &       &       &       &  \\
      & MillerCoors & 0.577*** & 0.262* & 0.423*** &       &       &  \\
      &       & (0.062) & (0.131) & (0.145) &       &       &  \\
      & Other Vendors & 0.490*** & 0.375*** & 0.208** & 0.335*** &       &  \\
      &       & (0.050) & (0.098) & (0.102) & (0.045) &       &  \\
      & Malt Beverage & 0.511*** & 0.271** & 0.473*** & 0.575*** & 0.759*** &  \\
      &       & (0.082) & (0.114) & (0.155) & (0.073) & (0.059) &  \bigstrut[b]\\
\hline

\multicolumn{8}{p{1.15\textwidth}}{Notes:
This table presents the posterior means and standard deviations (in parentheses) of the model parameters from the main specification, introduced in Subsection \ref{sec:data_and_specification}, and by the benchmark model, the time-invariant random-effects (RE) bundle demand model without endogenous regressors (Exo).
*** $p<0.01$, ** $p<0.05$, * $p<0.1$. The significance levels are based on the posterior credibility intervals.
The upper panel details the good utility equations $\bar{u}_{ijt}$ for the six goods.
The six columns correspond to the six goods.
Note that the price coefficient $\alpha$ is common across all six utility equations.
The bottom panel details the bundle effects $\Gamma_{i(j_1,j_2)t}$ for all fifteen pairs of goods.
The coefficient on (demeaned) family size is assumed to be homogeneous across all pairs of bundles.
} \bigstrut[t]\\
\end{tabular}}
	\end{table}
	
	\begin{table}[!h]
		\centering
		\caption{Estimates for the Exogenous TI-FA Bundle Demand Model}
		\resizebox{\textwidth}{!}{
			\footnotesize
			\begin{tabular}{llcccccc}
\hline
      & \multicolumn{1}{r}{$j = $} & 1     & 2     & 3     & 4     & 5     & 6 \bigstrut[t]\\
      &       & AB InBev & Boston Beer & Heineken & MillerCoors & Other Vendors & Malt Beverage \bigstrut[b]\\
\hline
\multicolumn{8}{l}{\boldmath{}\textbf{Good Utility $\bar{u}_{ijt}$}\unboldmath{}} \bigstrut[t]\\
      & Price & \multicolumn{6}{c}{-0.230***} \\
      &       & \multicolumn{6}{c}{(0.073)} \\
      & Constant & -2.677*** & -4.327*** & -5.096*** & -1.908*** & -2.555*** & -1.807*** \\
      &       & (0.622) & (0.707) & (0.814) & (0.586) & (0.568) & (0.653) \\
      & No. of Shopping Trips & 0.040*** & 0.020*** & -0.002 & 0.047*** & 0.055*** & 0.004 \\
      &       & (0.005) & (0.006) & (0.006) & (0.005) & (0.004) & (0.005) \\
      & Household Income & 0.066 & 0.173*** & 0.234*** & 0.039 & 0.152*** & 0.030 \\
      &       & (0.050) & (0.043) & (0.056) & (0.048) & (0.044) & (0.040) \\
      & Cabled TV & -0.146 & 0.204* & 0.329** & 0.231** & 0.112 & 0.141 \\
      &       & (0.109) & (0.121) & (0.147) & (0.106) & (0.087) & (0.090) \\
      & Pittsfield & 0.499*** & 1.035*** & 0.757*** & -0.734*** & -0.350*** & 0.402*** \\
      &       & (0.115) & (0.096) & (0.109) & (0.119) & (0.132) & (0.083) \bigstrut[b]\\
\hline
\multicolumn{8}{l}{\boldmath{}\textbf{Bundle-Effects $\Gamma_{i(j_1,j_2)t}$}\unboldmath{}} \bigstrut[t]\\
      & Family Size (Demeaned) & \multicolumn{6}{c}{0.016***} \\
      &       & \multicolumn{6}{c}{(0.007)} \\
      & Boston Beer & 0.939*** &       &       &       &       &  \\
      &       & (0.121) &       &       &       &       &  \\
      & Heineken USA & 0.495*** & -0.103 &       &       &       &  \\
      &       & (0.139) & (0.108) &       &       &       &  \\
      & MillerCoors & 0.648*** & 0.179* & 1.375*** &       &       &  \\
      &       & (0.061) & (0.103) & (0.157) &       &       &  \\
      & Other Vendors & 0.448*** & 0.256*** & 0.375** & 0.344*** &       &  \\
      &       & (0.057) & (0.090) & (0.133) & (0.044) &       &  \\
      & Malt Beverage & 0.739*** & 1.050*** & 0.922*** & 0.713*** & 0.714*** &  \\
      &       & (0.086) & (0.119) & (0.174) & (0.077) & (0.067) &  \bigstrut[b]\\
\hline
\multicolumn{8}{p{1.15\textwidth}}{Notes:
This table presents the posterior means and standard deviations (in parentheses) of the model parameters from the main specification, introduced in Subsection \ref{sec:data_and_specification}, and by the benchmark model, the time-invariant factor-augmented (TI-FA) bundle demand model without endogenous regressors (Exo).
*** $p<0.01$, ** $p<0.05$, * $p<0.1$. The significance levels are based on the posterior credibility intervals.
The upper panel details the good utility equations $\bar{u}_{ijt}$ for the six goods.
The six columns correspond to the six goods.
Note that the price coefficient $\alpha$ is common across all six utility equations.
The bottom panel details the bundle effects $\Gamma_{i(j_1,j_2)t}$ for all fifteen pairs of goods.
The coefficient on (demeaned) family size is assumed to be homogeneous across all pairs of bundles.
} \bigstrut[t]\\
\end{tabular}}
	\end{table}
	
	\begin{table}[!h]
		\centering
		\caption{Estimates for the Endogenous TI-FA Bundle Demand Model}
		\resizebox{\textwidth}{!}{
			\footnotesize
			\begin{tabular}{llcccccc}
\hline
      & \multicolumn{1}{r}{$j = $} & 1     & 2     & 3     & 4     & 5     & 6 \bigstrut[t]\\
      &       & AB InBev & Boston Beer & Heineken & MillerCoors & Other Vendors & Malt Beverage \bigstrut[b]\\
\hline
\multicolumn{8}{l}{\boldmath{}\textbf{Good Utility $\bar{u}_{ijt}$}\unboldmath{}} \bigstrut[t]\\
      & Price & \multicolumn{6}{c}{-0.249***} \\
      &       & \multicolumn{6}{c}{(0.073)} \\
      & Constant & -2.719*** & -5.308*** & -7.029*** & -1.665*** & -2.498*** & -1.896** \\
      &       & (0.724) & (0.988) & (0.929) & (0.652) & (0.622) & (0.764) \\
      & No. of Shopping Trips & 0.046*** & 0.036*** & 0.027*** & 0.054*** & 0.055*** & 0.019*** \\
      &       & (0.005) & (0.008) & (0.011) & (0.005) & (0.004) & (0.006) \\
      & Household Income & 0.072 & 0.226*** & 0.327*** & 0.013 & 0.153*** & 0.028 \\
      &       & (0.061) & (0.067) & (0.067) & (0.052) & (0.046) & (0.052) \\
      & Cabled TV & -0.157 & 0.159 & 0.338 & 0.220* & 0.119 & 0.095 \\
      &       & (0.124) & (0.192) & (0.232) & (0.122) & (0.089) & (0.108) \\
      & Pittsfield & 0.540*** & 1.187*** & 0.854*** & -0.808*** & -0.351** & 0.426*** \\
      &       & (0.101) & (0.135) & (0.136) & (0.115) & (0.135) & (0.094) \bigstrut[b]\\
\hline
\multicolumn{8}{l}{\boldmath{}\textbf{Bundle-Effects $\Gamma_{i(j_1,j_2)t}$}\unboldmath{}} \bigstrut[t]\\
      & Family Size (Demeaned) & \multicolumn{6}{c}{0.019**} \\
      &       & \multicolumn{6}{c}{(0.010)} \\
      & Boston Beer & 0.606*** &       &       &       &       &  \\
      &       & (0.112) &       &       &       &       &  \\
      & Heineken USA & 0.320*** & 0.454*** &       &       &       &  \\
      &       & (0.129) & (0.159) &       &       &       &  \\
      & MillerCoors & 0.577*** & 0.200* & 0.414*** &       &       &  \\
      &       & (0.060) & (0.114) & (0.118) &       &       &  \\
      & Other Vendors & 0.504*** & 0.383*** & 0.191* & 0.342*** &       &  \\
      &       & (0.054) & (0.084) & (0.101) & (0.055) &       &  \\
      & Malt Beverage & 0.510*** & 0.275** & 0.467*** & 0.566*** & 0.758*** &  \\
      &       & (0.073) & (0.114) & (0.129) & (0.072) & (0.061) &  \bigstrut[b]\\
\hline
\multicolumn{8}{p{1.15\textwidth}}{Notes:
This table presents the posterior means and standard deviations (in parentheses) of the model parameters from the main specification, introduced in Subsection \ref{sec:data_and_specification}, and by the benchmark model, the time-invariant factor-augmented (TI-FI) bundle demand model with endogenous regressors (Endo).
*** $p<0.01$, ** $p<0.05$, * $p<0.1$. The significance levels are based on the posterior credibility intervals.
The upper panel details the good utility equations $\bar{u}_{ijt}$ for the six goods.
The six columns correspond to the six goods.
Note that the price coefficient $\alpha$ is common across all six utility equations.
The bottom panel details the bundle effects $\Gamma_{i(j_1,j_2)t}$ for all fifteen pairs of goods.
The coefficient on (demeaned) family size is assumed to be homogeneous across all pairs of bundles.
} \bigstrut[t]\\
\end{tabular}}
	\end{table}
	
	\begin{table}[!h]
		\centering
		\caption{Estimates for the Exogenous TV-FA Bundle Demand Model}
		\resizebox{\textwidth}{!}{
			\footnotesize
			\begin{tabular}{llcccccc}
\hline
      & \multicolumn{1}{r}{$j = $} & 1     & 2     & 3     & 4     & 5     & 6 \bigstrut[t]\\
      &       & AB InBev & Boston Beer & Heineken & MillerCoors & Other Vendors & Malt Beverage \bigstrut[b]\\
\hline
\multicolumn{8}{l}{\boldmath{}\textbf{Good Utility $\bar{u}_{ijt}$}\unboldmath{}} \bigstrut[t]\\
      & Price & \multicolumn{6}{c}{-0.280***} \\
      &       & \multicolumn{6}{c}{(0.097)} \\
      & Constant & -2.820*** & -5.636*** & -6.961*** & -1.557** & -2.049*** & -2.227** \\
      &       & (0.681) & (1.061) & (1.202) & (0.668) & (0.696) & (0.882) \\
      & No. of Shopping Trips & 0.046*** & 0.032*** & 0.030*** & 0.054*** & 0.053*** & 0.022*** \\
      &       & (0.006) & (0.008) & (0.009) & (0.005) & (0.005) & (0.006) \\
      & Household Income & 0.085* & 0.237*** & 0.327*** & 0.002 & 0.120*** & 0.053 \\
      &       & (0.050) & (0.073) & (0.089) & (0.053) & (0.046) & (0.053) \\
      & Cabled TV & -0.191 & 0.210 & 0.349* & 0.301*** & 0.116 & 0.057 \\
      &       & (0.116) & (0.152) & (0.203) & (0.100) & (0.099) & (0.125) \\
      & Pittsfield & 0.502*** & 1.268*** & 0.873*** & -0.799*** & -0.304* & 0.458*** \\
      &       & (0.125) & (0.138) & (0.122) & (0.130) & (0.154) & (0.111) \bigstrut[b]\\
\hline
\multicolumn{8}{l}{\boldmath{}\textbf{Bundle-Effects $\Gamma_{i(j_1,j_2)t}$}\unboldmath{}} \bigstrut[t]\\
      & Family Size (Demeaned) & \multicolumn{6}{c}{0.022*} \\
      &       & \multicolumn{6}{c}{(0.012)} \\
      & Boston Beer & 0.656*** &       &       &       &       &  \\
      &       & (0.118) &       &       &       &       &  \\
      & Heineken USA & 0.176 & 0.281 &       &       &       &  \\
      &       & (0.144) & (0.166) &       &       &       &  \\
      & MillerCoors & 0.575*** & 0.235* & 0.480*** &       &       &  \\
      &       & (0.067) & (0.127) & (0.122) &       &       &  \\
      & Other Vendors & 0.487*** & 0.410*** & 0.277** & 0.332*** &       &  \\
      &       & (0.058) & (0.099) & (0.116) & (0.051) &       &  \\
      & Malt Beverage & 0.451*** & 0.342*** & 0.439*** & 0.606*** & 0.673*** &  \\
      &       & (0.084) & (0.147) & (0.147) & (0.073) & (0.070) &  \bigstrut[b]\\
\hline
\multicolumn{8}{p{1.15\textwidth}}{Notes:
This table presents the posterior means and standard deviations (in parentheses) of the model parameters from the main specification, introduced in Subsection \ref{sec:data_and_specification}, and by the benchmark model, the time-varying factor-augmented (TV-FA) bundle demand model without endogenous regressors (Exo).
*** $p<0.01$, ** $p<0.05$, * $p<0.1$. The significance levels are based on the posterior credibility intervals.
The upper panel details the good utility equations $\bar{u}_{ijt}$ for the six goods.
The six columns correspond to the six goods.
Note that the price coefficient $\alpha$ is common across all six utility equations.
The bottom panel details the bundle effects $\Gamma_{i(j_1,j_2)t}$ for all fifteen pairs of goods.
The coefficient on (demeaned) family size is assumed to be homogeneous across all pairs of bundles.
} \bigstrut[t]\\
\end{tabular}}
	\end{table}
	
	\clearpage
	\subsection{Price Elasticity Estimates from the Benchmark Models}
	\label{sec:appendix_sup_elasticities}
	
	\begin{table}[!h]
		\centering
		\caption{Estimates of Own- and Cross-Price Elasticities by the Exogenous RE Bundle Demand Model}
		\resizebox{\textwidth}{!}{
			\footnotesize
			\begin{tabular}{llcccccc}
\hline
\multicolumn{2}{r}{Good $k=$} & 1     & 2     & 3     & 4     & 5     & 6 \bigstrut[t]\\
\multicolumn{1}{l}{Price $j=$} &       & AB InBev & Boston Beer & Heineken & MillerCoors & Other Vendors & Malt Beverage \bigstrut[b]\\
\hline
\multicolumn{1}{l}{1} & \multicolumn{1}{l}{AB InBev} & -1.053*** & -0.050 & -0.007 & -0.033 & -0.028 & -0.038 \bigstrut[t]\\
      &       & (0.336) & (0.077) & (0.100) & (0.025) & (0.023) & (0.050) \\
\multicolumn{1}{l}{2} & \multicolumn{1}{l}{Boston Beer} & -0.020 & -2.223*** & -0.026 & -0.001 & -0.011 & 0.005 \\
      &       & (0.026) & (0.726) & (0.097) & (0.016) & (0.018) & (0.034) \\
\multicolumn{1}{l}{3} & \multicolumn{1}{l}{Heineken} & -0.002 & -0.016 & -2.244*** & -0.005 & 0.001 & -0.009 \\
      &       & (0.021) & (0.060) & (0.748) & (0.013) & (0.013) & (0.030) \\
\multicolumn{1}{l}{4} & \multicolumn{1}{l}{MillerCoors} & -0.046 & -0.002 & -0.026 & -0.785*** & -0.018 & -0.054 \\
      &       & (0.036) & (0.072) & (0.093) & (0.250) & (0.023) & (0.052) \\
\multicolumn{1}{l}{5} & \multicolumn{1}{l}{Other Vendors} & -0.063 & -0.059 & 0.006 & -0.029 & -1.082*** & -0.136** \\
      &       & (0.048) & (0.111) & (0.129) & (0.033) & (0.340) & (0.079) \\
\multicolumn{1}{l}{6} & \multicolumn{1}{l}{Malt Beverage} & -0.035 & 0.011 & -0.032 & -0.036 & -0.056** & -2.116*** \\
      &       & (0.037) & (0.076) & (0.106) & (0.027) & (0.030) & (0.675) \bigstrut[b]\\
\hline
\multicolumn{8}{p{1.12\textwidth}}{Notes:
This table presents the estimated price elasticities by the main model, the time-invariant random-effects (RE) bundle demand model without endogenous regressors (Exo).
Cell entries $(j,k)$, where $j$ indexes row and $k$ indexes column, give the posterior mean (and standard deviation in parentheses) of the percentage change in the probability of using $k$ substance in response to a percentage increase in the price of substance $j$.
*** $p<0.01$, ** $p<0.05$, * $p<0.1$.
The significance levels are based on the posterior credibility intervals.
} \bigstrut[t]\\
\end{tabular}}
	\end{table}
	
	\begin{table}[!h]
		\centering
		\caption{Estimates of Own- and Cross-Price Elasticities by the Exogenous TI-FA Bundle Demand Model}
		\resizebox{\textwidth}{!}{
			\footnotesize
			\begin{tabular}{llcccccc}
\hline
\multicolumn{2}{r}{Good $k=$} & 1     & 2     & 3     & 4     & 5     & 6 \bigstrut[t]\\
\multicolumn{1}{l}{Price $j=$} &       & AB InBev & Boston Beer & Heineken & MillerCoors & Other Vendors & Malt Beverage \bigstrut[b]\\
\hline
\multicolumn{1}{l}{1} & \multicolumn{1}{l}{AB InBev} & -1.003*** & -0.109 & -0.050 & -0.037* & -0.024 & -0.071 \bigstrut[t]\\
      &       & (0.329) & (0.086) & (0.111) & (0.025) & (0.022) & (0.053) \\
\multicolumn{1}{l}{2} & \multicolumn{1}{l}{Boston Beer} & -0.044* & -2.358*** & 0.116 & 0.006 & -0.004 & -0.046 \\
      &       & (0.030) & (0.787) & (0.132) & (0.017) & (0.018) & (0.039) \\
\multicolumn{1}{l}{3} & \multicolumn{1}{l}{Heineken} & -0.012 & 0.071 & -2.629*** & -0.031** & -0.005 & -0.034 \\
      &       & (0.023) & (0.081) & (0.894) & (0.018) & (0.015) & (0.037) \\
\multicolumn{1}{l}{4} & \multicolumn{1}{l}{MillerCoors} & -0.052 & 0.018 & -0.169** & -0.762*** & -0.018 & -0.071 \\
      &       & (0.037) & (0.074) & (0.113) & (0.248) & (0.023) & (0.055) \\
\multicolumn{1}{l}{5} & \multicolumn{1}{l}{Other Vendors} & -0.055 & -0.025 & -0.046 & -0.030 & -1.025*** & -0.126* \\
      &       & (0.047) & (0.113) & (0.146) & (0.033) & (0.331) & (0.077) \\
\multicolumn{1}{l}{6} & \multicolumn{1}{l}{Malt Beverage} & -0.065* & -0.104 & -0.124 & -0.047* & -0.052** & -2.111*** \\
      &       & (0.042) & (0.086) & (0.132) & (0.030) & (0.029) & (0.686) \bigstrut[b]\\
\hline
\multicolumn{8}{p{1.12\textwidth}}{Notes:
This table presents the estimated price elasticities by the main model, the time-invariant factor-augmented (TI-FA) bundle demand model without endogenous regressors (Exo).
Cell entries $(j,k)$, where $j$ indexes row and $k$ indexes column, give the posterior mean (and standard deviation in parentheses) of the percentage change in the probability of using $k$ substance in response to a percentage increase in the price of substance $j$.
*** $p<0.01$, ** $p<0.05$, * $p<0.1$.
The significance levels are based on the posterior credibility intervals.
} \bigstrut[t]\\
\end{tabular}}
	\end{table}
	
	\begin{sidewaystable}[!h]
		\centering
		\caption{Product Baskets for Construction of Individual-Level Prices}
		\resizebox{\textwidth}{!}{
			\footnotesize
\begin{tabular}{llllcccccccc}
\hline
$j$   & \multicolumn{3}{l}{Good} & \multicolumn{5}{c}{Number of}         &       & \multicolumn{2}{c}{Price Basket} \bigstrut\\
\cline{5-9}\cline{11-12}      &       & \multicolumn{2}{l}{Category (Level 1)} & Parent Companies & Vendors & Brands & Products & Package Sizes &       & \multicolumn{2}{l}{Definition} \bigstrut[t]\\
      &       &       & Small Category (Level 2) &  (Level 3) & (Level 4) & (Level 5) & (UPC) & (Eq. Volumes) &       &       & Number of Items \bigstrut[b]\\
\hline
1     & \multicolumn{3}{l}{\textbf{Sugary Soft Drinks}} &       &       &       &       &       &       & \multicolumn{2}{l}{\textbf{L1-L2-L3-L4-L5-Volume}} \bigstrut[t]\\
      &       & \multicolumn{2}{l}{Carbonated Beverages} &       &       &       &       &       &       &       & \textbf{292} \\
      &       &       & Regular Soft Drinks & \textbf{35} & \textbf{36} & \textbf{97} & \textbf{636} & \textbf{18} &       &       &  \bigstrut[b]\\
\hline
2     & \multicolumn{3}{l}{\textbf{Diet Soft Drinks}} &       &       &       &       &       &       & \multicolumn{2}{l}{\textbf{L1-L2-L3-L4-L5-Volume}} \bigstrut[t]\\
      &       & \multicolumn{2}{l}{Carbonated Beverages} &       &       &       &       &       &       &       & \textbf{228} \\
      &       &       & Low Calorie Soft Drinks & \textbf{15} & \textbf{16} & \textbf{75} & \textbf{354} & \textbf{17} &       &       &  \bigstrut[b]\\
\hline
3     & \multicolumn{3}{l}{\textbf{Carbonated Water}} &       &       &       &       &       &       & \multicolumn{2}{l}{\textbf{L1-L2-L3-L4-L5-Volume}} \bigstrut[t]\\
      &       & \multicolumn{2}{l}{Carbonated Beverages} &       &       &       &       &       &       &       & \textbf{32} \\
      &       &       & Seltzer, Tonic Water, Club Soda & \textbf{8} & \textbf{8} & \textbf{12} & \textbf{209} & \textbf{10} &       &       &  \bigstrut[b]\\
\hline
4     & \multicolumn{3}{l}{\textbf{Milk Drinks}} & \textbf{19} & \textbf{23} & \textbf{39} & \textbf{158} & \textbf{10} &       & \multicolumn{2}{l}{\textbf{L1-L2-L3-L4-Volume}} \bigstrut[t]\\
      &       & \multicolumn{2}{l}{Milk} &       &       &       &       &       &       &       & \textbf{50} \\
      &       &       & Flavored Milk, Eggnog, Buttermilk & 14    & 17    & 33    & 142   & 6     &       &       &  \\
      &       &       & Milkshakes, Non-Dairy Drinks & 5     & 6     & 6     & 16    & 4     &       &       &  \bigstrut[b]\\
\hline
5     & \multicolumn{3}{l}{\textbf{Salty Snacks}} & \textbf{210} & \textbf{215} & \textbf{349} & \textbf{1708} & \textbf{196} &       & \multicolumn{2}{l}{\textbf{L1-L2-L3-L4-Volume}} \bigstrut[t]\\
      &       & \multicolumn{2}{l}{Salty Snacks} &       &       &       &       &       &       &       & \textbf{642} \\
      &       &       & Cheese Snacks & 21    & 21    & 29    & 125   & 27    &       &       &  \\
      &       &       & Corn Snacks (No Tortilla Chips) & 12    & 12    & 18    & 57    & 16    &       &       &  \\
      &       &       & Other Salted Snacks & 53    & 55    & 84    & 290   & 35    &       &       &  \\
      &       &       & Pork Rinds & 10    & 10    & 10    & 25    & 4     &       &       &  \\
      &       &       & Potato Chips & 25    & 27    & 62    & 565   & 29    &       &       &  \\
      &       &       & Pretzels & 22    & 22    & 28    & 194   & 29    &       &       &  \\
      &       &       & Ready-to-Eat Popcorn, Caramel Corn & 33    & 33    & 59    & 165   & 25    &       &       &  \\
      &       &       & Tortilla, Tostada Chips & 34    & 35    & 59    & 287   & 31    &       &       &  \bigstrut[b]\\
\hline
\end{tabular}
}
		\label{tab:soda_price_construction}
	\end{sidewaystable}
	
\end{document}